\documentclass[a4paper,fleqn,usenatbib]{mnras}

\usepackage[T1]{fontenc}
\usepackage{ae,aecompl}

\usepackage{graphicx}	
\usepackage{amsmath}	
\usepackage{amssymb}	
\usepackage{array}
\usepackage{booktabs}
\usepackage{pdflscape}
\newcommand{\msun}{\mathrm{M}_{\sun}}
\newcommand{\lsun}{\mathrm{L}_{\sun}}
\newcommand{\ssd}{\hat{\sigma}}
\newcommand{\mni}{M_{\mathrm{Ni}}}
\newcommand{\loi}{L_\mathrm{[\ion{O}{I}]}^{350\text{d}}}
\newcommand{\mv}{M_V^{50\text{d}}}
\newcommand{\bv}{{B\!-\!V}}
\newcommand{\Lf}{L_\text{prog}}

\everycr={\noalign{\global\advance\stepno by 1}}%

\title[Luminosity distribution of SN II progenitors]{Luminosity distribution of Type II supernova progenitors}
\author[\'{O} Rodr\'{i}guez]{\'{O}smar Rodr\'{i}guez$^{1}$\thanks{E-mail: olrodrig@gmail.com} 
\\
$^{1}$School of Physics and Astronomy, Tel Aviv University, Tel Aviv 69978, Israel
}

\date{Accepted XXX. Received YYY; in original form ZZZ}

\pubyear{2022}

\begin{document}
\label{firstpage}
\pagerange{\pageref{firstpage}--\pageref{lastpage}}
\maketitle

\begin{abstract} 
I present progenitor luminosities ($L$) for a sample of 112 Type II supernovae (SNe~II), computed directly from progenitor photometry and the bolometric correction technique, or indirectly from empirical correlations between progenitor luminosity and [\ion{O}{I}]\,$\lambda\lambda$6300,~6364 line luminosity at 350\,d since explosion, $^{56}$Ni mass, or absolute $V$-band magnitude at 50\,d since explosion. To calibrate these correlations, I use twelve SNe~II with progenitor luminosities measured from progenitor photometry. I find that the correlations mentioned above are strong and statistically significant, and allow to estimate progenitor luminosities to a precision between 20 and 24~per~cent. I correct the SN sample for selection bias and define a subsample of 112~SNe~II with progenitor luminosities between ${\log(L/L_{\sun})=4.6}$\,dex, corresponding to the completeness limit of the corrected sample, and the maximum observed progenitor luminosity of $\log(L/L_{\sun})=5.091$\,dex. The luminosity distribution for this subsample is statistically consistent with those for red supergiants (RSGs) in LMC, SMC, M31, and M33 with $4.6\leq\log(L/L_{\sun})\leq5.091$. This supports that SN~II progenitors correspond to RSGs. The conspicuous absence of SN~II progenitors with $\log(L/L_{\sun})>5.1$\,dex with respect to what is observed in RSG luminosity distributions, known as the RSG problem, is significant at a $5.2\pm0.5\,\sigma$ level.
\end{abstract}




\section{Introduction}\label{sec:introduction}
Type II supernovae \citep[SNe~II;][]{1941PASP...53..224M} are the explosions of massive stars with an important amount of hydrogen in their envelope at the moment of explosion. Thus, the classification of an SN as a Type II is based on the presence of H lines in its spectrum. Among SNe~II there are some objects showing narrow H emission lines in the spectra, indicative of interaction of the ejecta with circumstellar material \citep[SNe~IIn;][]{1990MNRAS.244..269S},\footnote{This group includes SNe~IIn/II and LLEV SNe~II, described in \citet{2020MNRAS.494.5882R}.} SNe having long-rising light curves similar to SN 1987A \citep[e.g.][]{1988AJ.....95...63H,2016AA...588A...5T}, and some SNe showing peculiar characteristics that make them unique (e.g. OGLE14-073, \citealt{2017NatAs...1..713T}; iPTF14hls, \citealt{2017Natur.551..210A}; ASASSN-15nx, \citealt{2018ApJ...862..107B}). The rest of events account for about 90 per cent of the SN~II population in a volume-limited sample \citep[e.g.][]{2017PASP..129e4201S}. SNe belonging to the SN~IIn and long-rising SN subgroups, and those with peculiar characteristics are not included in the present analysis.

Progenitors of SNe~II have been directly identified on pre-explosion images \citep[e.g.][]{2009ARAA..47...63S,2015PASA...32...16S,2017RSPTA.37560277V}, some of them being confirmed as such by their disappearance in late-time high-resolution images. The spectral energy distributions (SEDs) and colour indices of SN~II progenitors fit well with those of red supergiant (RSG) stars \citep[e.g.][]{2003PASP..115.1289V,2019ApJ...875..136V,2004Sci...303..499S,2013MNRAS.431L.102M,2019AA...622L...1O,2021AA...645L...7O}. The identification of RSGs as SN~II progenitors is consistent with results from pioneering works \citep[e.g.][]{1971ApSS..10...28G,1976ApJ...207..872C,1977ApJS...33..515F}, who found that SN~II progenitors are stars with large radii and massive H-rich envelopes.

Another important parameter characterizing SN progenitors is the luminosity ($L$), which is computed from pre-explosion photometry using the bolometric correction (BC) technique or the SED integration. Since progenitors are observed in the final stage of their evolution (a few years before the SN explosion), the observed progenitor luminosity ($\Lf$) is also called final luminosity. The progenitor luminosity, along with an initial mass-final luminosity relation from stellar evolution models, is used to determine the progenitor initial mass $M_\text{i}$ (e.g. \citealt{2009MNRAS.395.1409S,2015PASA...32...16S}; and references therein). With the increase in the number of detected progenitors, it became possible to infer properties of the population of RSGs that explode as SNe~II. Using a sample of 20~SN~II progenitors and assuming a Salpeter initial mass function (IMF), \citet{2009MNRAS.395.1409S} derived a maximum initial mass of $16.5\pm1.5\,\msun$, which is systematically lower than the maximum RSG mass of around $25\,\msun$. In particular, the authors found that the lack of SN~II~progenitors with $M_\text{i}$ between $17$ and  $25\,\msun$ with respect to what is expected for a Salpeter IMF is statistically significant at $2.4\,\sigma$ confidence. They termed this discrepancy the ``RSG problem''. Later studies, which included new and updated progenitor luminosities \citep[e.g.][]{2015PASA...32...16S,2018MNRAS.474.2116D,2020MNRAS.493..468D}, increased the maximum initial mass to 18--$19\,\msun$\footnote{\citet{2014MNRAS.440.1917D} reported a similar upper limit of $\sim19\,\msun$ based on the lack of SNe~II with mass-loss rate $>10^{-5}\,\msun$\,yr$^{-1}$.} and set the statistical significance of the RSG problem to around $2\,\sigma$. Based on the analysis of 24 SN~II progenitors, \citet{2020MNRAS.493..468D} concluded that it is necessary to at least double the sample size to determine whether the RSG problem is statistically significant.

Alternative methods to measure $M_\text{i}$ appear as promising tools to increase the number of SNe~II with $M_\text{i}$ estimates. One of these methods is the age-dating technique \citep[e.g.][]{2004ApJ...615L.113M,2011ApJ...742L...4M}, where the age of the stellar population in the SN vicinity is adopted as the age of the progenitor, which allows to estimate its initial mass. This technique has been applied to small SN samples \citep[e.g.][]{2014ApJ...791..105W,2018ApJ...860...39W,2017MNRAS.469.2202M,2021MNRAS.506..781D}. Another method to estimate $M_\text{i}$ is by fitting hydrodynamical models to SN light curves \citep[e.g.][]{1998ApJ...496..454B,2004AstL...30..293U,2011ApJ...729...61B,2011ApJ...741...41P,2015ApJ...814...63M}. This method has been used to study individual SNe and small SN samples \citep[e.g.][]{2018ApJ...858...15M,2018NatAs...2..808F,2019PASA...36...41E,2019ApJ...880...59R,2019AA...629A.124M,2020AA...642A.143M,2021MNRAS.505..116U}. Recently, \citet{2022AA...660A..41M} presented results from hydrodynamical modelling to 53~SNe~II, finding a maximum initial mass of $21.3\,\msun$. A third alternative method to infer $M_\text{i}$ is by comparing late-time spectra with nebular spectra models \citep[e.g.][]{2012AA...546A..28J,2014MNRAS.439.3694J,2018MNRAS.475..277J,2021AA...652A..64D}. In particular, the luminosity of the [\ion{O}{I}]\,$\lambda\lambda$6300,~6364 doublet line ($L_{[\ion{O}{I}]}$) has been shown as a promising observable to estimate $M_\text{i}$ \citep[e.g.][]{2014MNRAS.439.3694J,2015MNRAS.448.2482J}, while nebular spectra models of \citet{2021AA...652A..64D} show a dependence of $L_{[\ion{O}{I}]}$ at 350\,d since explosion ($\loi$) on initial mass. This relation arises because $L_{[\ion{O}{I}]}$ depends on $^{56}$Ni mass ($\mni$) and oxygen mass \citep[e.g.][]{2003MNRAS.338..939E,2011AcA....61..179E}, where $\mni$ correlates with $M_\text{i}$ \citep[e.g.][]{2012ApJ...744...26O} while oxygen mass depends on the helium-core mass, which in turns depends on $M_\text{i}$ \citep[e.g.][]{1995ApJS..101..181W}. The method of estimating initial mass by comparing late-time spectra with spectral models has been applied to a small number of SNe \citep[e.g.][]{2012AA...546A..28J,2014MNRAS.439.3694J,2015MNRAS.448.2482J,2018MNRAS.475..277J,2017MNRAS.467..369S,2021AA...652A..64D}.

In general, methods to infer $M_\text{i}$ require assuming a stellar evolution model, which depends not only on initial mass but also on composition, convection, rotation, mass-loss, binary interaction, among others \citep[e.g.][]{2004MNRAS.353...87E,2015AA...575A..60M,2018ApJS..237...13L,2019ApJ...881..158S,2021AA...645A...6Z}. Initial mass values computed with the methods mentioned earlier are, therefore, affected by systematic errors related to the stellar evolution modelling and ignorance of the progenitor properties. \citet{2020MNRAS.493..468D} showed that the SN~II progenitor population can be studied in terms of progenitor luminosity instead of initial mass, thus preventing adding the systematic uncertainties mentioned above. In that work, the authors compared the luminosity distribution for SN~II progenitors to that observed for RSGs in LMC. This kind of comparison allows to identify similarities and differences between both populations in a completely empirical way. For example, the RSG problem can be reformulated as the lack of SN~II~progenitors with luminosity greater than $\log(L/\lsun)=5.1$\,dex \citep[e.g.][]{2015PASA...32...16S} with respect to what is \textit{observed} in RSG luminosity distributions. On the other hand, the disadvantage of using $\Lf$ is that analysis of the luminosity distribution for SN~II progenitors is restricted to the small sample of SNe with available progenitor photometry.

An alternative method to increase the number of SNe~II with $\Lf$ measurements is by inferring $\Lf$ indirectly from empirical correlations. \citet{2011MNRAS.417.1417F} found a relation between $\Lf$ and $\mni$, which was also reported by \citet{2015arXiv150602655K}. Unfortunately, the authors did not report the strength, significance, or the analytical expression for the observed correlation. Recently, \citet{2018MNRAS.474.2116D,2020MNRAS.493..468D} have presented an updated list of SN~II progenitors and their luminosities, while updated $\mni$ estimates for many of those SNe were reported by \citet{2021MNRAS.505.1742R}. Therefore, it is possible to analyse the correlation between $\Lf$ and $\mni$ with new and improved data. A few other works have analysed empirically the dependence of SN~II observables on initial mass computed from $\Lf$ \citep[e.g.][]{2009MNRAS.395.1409S,2012ApJ...744...26O,2012MNRAS.420.3451M,2013MNRAS.436.3224P}. In particular, \citet{2013MNRAS.436.3224P} suggested a correlation between $M_\text{i}$ and expansion velocity of the photosphere at 50\,d since explosion ($v_\text{50d}$). Because of the correlation between $v_\text{50d}$ and the absolute $V$-band magnitude at 50\,d since explosion ($\mv$) observed for SNe~II \citep{2003ApJ...582..905H}, the relation suggested by \citet{2013MNRAS.436.3224P} could translate into a correlation between $\Lf$ and $\mv$. On the theoretical side, the relation between $\loi$ and $M_\text{i}$ shown by the nebular spectra models of \citet{2021AA...652A..64D} suggests a possible correlation between $\Lf$ and $\loi$.

In this work, I investigate empirical correlations between $\Lf$ and three SN observables: $\loi$, $\mni$, and $\mv$. I use these correlations to compute $\Lf$ values for 112~SNe~II collected from the literature. The aim is to construct the luminosity distribution for SN~II progenitors and compare it to observed RSG luminosity distributions.

The paper is organized as follows. In Section~2, I outline the relevant information on the data used in this study. In Section~3, I present methods to measure $\loi$, $\Lf$ from pre-explosion photometry, and to correct the SN sample for selection bias. In Section~4, I report the correlations between $\Lf$ and SN observables, the progenitor luminosity distribution, and the comparison with different RSG luminosity distributions. Comparison to previous work and discussion of systematics appear in Section~5. Conclusions are summarised in Section~6.

\section{Data Set}\label{sec:data_set}

\subsection{SN sample}
In this work I use the sample of 110~SNe~II analysed in \citet{2021MNRAS.505.1742R}. In that work, the authors collected SNe~II from the literature having photometry in the radioactive tail in at least one optical band ($V$, $r$, $R$, $i$, or $I$) with at least three photometric epochs between 95 and 320\,d since explosion. \citet{2021MNRAS.505.1742R} used these data to compute accurate $^{56}$Ni masses for the selected SNe. The authors also calculated distance moduli ($\mu$), explosion epochs ($t_\text{expl}$), host galaxy reddenings ($E_\bv^\text{host}$), $\mv$ values, and absolute $R$-band magnitudes at maximum ($M_R^\text{max}$, which are used to perform the correction for selection bias; see Section~\ref{sec:SBC}). Since the six quantities mentioned above were computed in an homogeneous way, the data presented in \citet{2021MNRAS.505.1742R} are suitable to carry out the present study. I also include SN~2018aoq, for which a progenitor candidate has been identified \citep{2019AA...622L...1O}, and SN~2015bs, which shows a prominent [\ion{O}{I}] doublet in its nebular spectrum \citep{2018NatAs...2..574A}. For these two SNe, I compute $\mu$, $t_\text{expl}$, $E_\bv^\text{host}$, $\mv$, $M_R^\text{max}$, and $^{56}$Ni mass in the same manner as in \citet{2021MNRAS.505.1742R} (see Appendix~\ref{sec:SNe_appendix}). The final sample of 112~SNe is listed in Table~\ref{table:SN_sample}. This includes the SN name (Column~1), the $^{56}$Ni mass (Column~2), $\mv$ (Column~3), and $M_R^\text{max}$ (Column~4).

\begin{table}
\caption{SN sample.}
\label{table:SN_sample}
\begin{tabular}{lccc}
\hline
 SN & $\log(\mni/\msun)$ & $\mv$ & $M_R^\text{max}$ \\
\hline
1980K       & $-1.462(147)$ & $-17.348(379)$  & $-18.764$ \\
1986I       & $-1.302(198)$ & $-16.911(590)$  & $-16.730$ \\
1988A       & $-1.109(185)$ & $-16.488(488)$  & $-16.639$ \\
1990E       & $-1.383(123)$ & $-16.702(341)$  & $-17.401$ \\
1990K       & $-1.470(118)$ & $-16.705(275)$  & $-17.784$ \\
1991G       & $-1.778(99)$  & $-15.365(277)$  & $-15.729$ \\
1991al      & $-1.629(68)$  & $-15.974(159)$  & $-16.836$ \\
1992H       & $-0.788(178)$ & $-17.670(413)$  & $-18.067$ \\
1992ba      & $-1.744(126)$ & $-15.733(305)$  & $-16.218$ \\
1994N       & $-2.283(106)$ & $-14.955(233)$  & $-15.371$ \\
1995ad      & $-1.230(115)$ & $-17.086(387)$  & $-17.782$ \\
1996W       & $-0.952(97)$  & $-17.469(256)$  & $-17.798$ \\
1997D       & $-2.064(144)$ & $<-14.436(423)$ & $-14.986$ \\
1999ca      & $-1.846(70)$  & $-16.804(167)$  & $-18.086$ \\
1999em      & $-1.296(62)$  & $-16.692(138)$  & $-17.192$ \\
1999ga      & $-1.446(120)$ & $<-16.693(285)$ & $-17.320$ \\
1999gi      & $-1.333(87)$  & $-16.250(207)$  & $-16.728$ \\
2001X       & $-1.395(97)$  & $-16.348(231)$  & $-16.726$ \\
2001dc      & $-2.119(110)$ & $-15.059(283)$  & $-15.283$ \\
2002gw      & $-1.631(103)$ & $-16.017(246)$  & $-16.310$ \\
2002hh      & $-1.082(69)$  & $-16.882(285)$  & $-17.107$ \\
2002hx      & $-1.186(64)$  & $-16.671(165)$  & $-17.596$ \\
2003B       & $-2.223(115)$ & $-14.768(271)$  & $-15.248$ \\
2003T       & $-1.344(101)$ & $-16.674(164)$  & $-17.131$ \\
2003Z       & $-2.262(116)$ & $-14.610(277)$  & $-14.958$ \\
2003fb      & $-1.482(111)$ & $-16.016(206)$  & $-16.582$ \\
2003gd      & $-1.694(99)$  & $<-16.405(149)$ & $-16.859$ \\
2003hd      & $-1.362(63)$  & $-17.056(156)$  & $-17.635$ \\
2003hk      & $-1.572(105)$ & $-17.101(211)$  & $-18.303$ \\
2003hn      & $-1.412(56)$  & $-16.819(136)$  & $-17.602$ \\
2003ho      & $-1.601(105)$ & $-16.554(267)$  & $-17.415$ \\
2003iq      & $-1.318(82)$  & $-16.763(200)$  & $-17.347$ \\
2004A       & $-1.604(116)$ & $-15.898(291)$  & $-16.159$ \\
2004dj      & $-1.902(130)$ & $-15.846(178)$  & $-16.232$ \\
2004eg      & $-2.126(188)$ & $<-15.077(551)$ & $-15.693$ \\
2004ej      & $-1.793(166)$ & $-16.494(380)$  & $-17.026$ \\
2004et      & $-1.037(71)$  & $-17.645(213)$  & $-17.974$ \\
2004fx      & $-1.802(173)$ & $-15.720(400)$  & $-15.984$ \\
2005af      & $-1.477(179)$ & $<-15.212(383)$ & $-17.413$ \\
2005au      & $-1.195(85)$  & $-17.291(273)$  & $-17.974$ \\
2005ay      & $-1.778(109)$ & $-15.512(239)$  & $-15.912$ \\
2005cs      & $-2.241(76)$  & $-15.371(133)$  & $-15.700$ \\
2005dx      & $-2.064(194)$ & $-15.583(445)$  & $-16.381$ \\
2006my      & $-1.674(141)$ & $<-15.361(473)$ & $-16.896$ \\
2006ov      & $-2.051(267)$ & $<-16.568(707)$ & $-16.195$ \\
2007aa      & $-1.522(143)$ & $-16.556(407)$  & $-16.703$ \\
2007hv      & $-1.326(125)$ & $-16.698(400)$  & $-17.158$ \\
2007it      & $-0.987(130)$ & $-17.400(417)$  & $-18.234$ \\
2007od      & --            & $-17.438(332)$  & $-18.027$ \\
2008K       & $-1.611(180)$ & $-16.806(378)$  & $-17.869$ \\
2008M       & $-1.605(172)$ & $-16.504(396)$  & $-17.236$ \\
2008aw      & $-1.074(114)$ & $-17.228(230)$  & $-18.181$ \\
2008bk      & $-2.064(94)$  & $-14.992(366)$  & $-15.245$ \\
2008gz      & $-1.253(81)$  & $<-16.366(194)$ & $-17.679$ \\
2008in      & $-1.640(258)$ & $-16.134(644)$  & $-16.596$ \\
2009N       & $-1.891(161)$ & $-15.357(404)$  & $-15.609$ \\
2009at      & $-1.750(115)$ & $-16.153(332)$  & $-17.283$ \\
2009ay      & $-0.932(116)$ & $-17.720(309)$  & $-18.552$ \\
2009bw      & $-1.737(120)$ & $-16.393(316)$  & $-17.309$ \\
2009dd      & $-1.474(112)$ & $-16.385(259)$  & $-16.993$ \\
2009hd      & $-1.959(66)$  & $-16.632(188)$  & $-17.327$ \\
2009ib      & $-1.356(75)$  & $-15.790(131)$  & $-16.266$ \\
2009md      & $-2.097(78)$  & $-15.365(195)$  & $-15.779$ \\
2010aj      & $-2.088(75)$  & $-16.764(203)$  & $-17.836$ \\

\end{tabular}
\end{table}
\begin{table}
\contcaption{}
\begin{tabular}{lccc}
\hline
 SN & $\log(\mni/\msun)$ & $\mv$ & $M_R^\text{max}$ \\
\hline
PTF10gva    & $-1.111(138)$ & --              & $-18.794$ \\
2011fd      & $-1.498(125)$ & --              & $-16.998$ \\
PTF11go     & $-1.593(160)$ & --              & $-16.743$ \\
PTF11htj    & $-1.309(175)$ & --              & $-16.946$ \\
PTF11izt    & $-1.651(159)$ & --              & $-16.348$ \\
2012A       & $-1.766(111)$ & $-16.454(284)$  & $-17.176$ \\
2012aw      & $-1.269(55)$  & $-16.847(135)$  & $-17.181$ \\
2012br      & $-1.219(254)$ & --              & $-17.764$ \\
2012cd      & $-1.038(168)$ & --              & $-18.680$ \\
2012ec      & $-1.545(115)$ & $-16.336(290)$  & $-16.695$ \\
PTF12grj    & $-1.565(169)$ & --              & $-16.963$ \\
PTF12hsx    & $-1.090(225)$ & --              & $-17.331$ \\
2013K       & $-1.593(123)$ & $-16.711(297)$  & $-16.430$ \\
2013ab      & $-1.482(95)$  & $-16.199(234)$  & $-16.804$ \\
2013am      & $-1.695(163)$ & $-15.470(358)$  & $-15.958$ \\
2013bu      & $-1.888(93)$  & $-16.104(217)$  & $-16.947$ \\
2013by      & $-1.512(151)$ & $-17.397(436)$  & $-18.411$ \\
2013ej      & $-1.551(60)$  & $-16.950(149)$  & $-17.900$ \\
2013fs      & $-1.323(106)$ & $-16.739(213)$  & $-17.623$ \\
2013hj      & $-1.092(73)$  & $-17.564(178)$  & $-18.270$ \\
iPTF13dkz   & $-1.173(205)$ & --              & $-16.470$ \\
LSQ13dpa    & $-1.101(135)$ & $-17.181(470)$  & $-17.694$ \\
2014G       & $-1.146(77)$  & $-17.203(199)$  & $-18.411$ \\
2014cx      & $-1.251(103)$ & $-16.668(263)$  & $-16.957$ \\
2014cy      & $-1.820(117)$ & $-15.584(269)$  & $-16.122$ \\
ASASSN-14dq & $-1.308(101)$ & $-16.967(249)$  & $-17.653$ \\
2014dw      & $-1.544(158)$ & $-16.238(487)$  & $-17.312$ \\
ASASSN-14ha & $-2.030(120)$ & $-15.963(369)$  & $-16.379$ \\
OGLE14-18   & $-1.367(123)$ & --              & $-17.103$ \\
2015V       & $-1.643(107)$ & $-15.751(287)$  & $-15.970$ \\
2015W       & $-1.372(110)$ & $-17.129(314)$  & $-17.893$ \\
2015an      & $-1.575(113)$ & $-17.184(363)$  & $-17.697$ \\
2015ba      & $-1.840(114)$ & $-17.020(221)$  & $-17.750$ \\
2015bs      & $-1.156(119)$ & $-16.988(325)$  & $-17.470$ \\
2015cz      & $-1.344(93)$  & $-17.158(272)$  & $-17.627$ \\
ASASSN-15oz & $-0.875(124)$ & $-17.682(316)$  & $-18.478$ \\
2016X       & $-1.349(147)$ & $-16.856(371)$  & $-17.588$ \\
2016aqf     & $-1.670(137)$ & $-16.024(394)$  & $-16.136$ \\
2016bkv     & $-1.817(106)$ & $-14.537(300)$  & $-15.741$ \\
2016gfy     & $-1.122(103)$ & $-17.145(263)$  & $-17.599$ \\
2016ija     & $-1.582(144)$ & $-16.691(490)$  & $-17.238$ \\
2017it      & $-1.257(60)$  & $-16.817(149)$  & $-17.178$ \\
2017ahn     & $-1.316(132)$ & $-16.751(464)$  & $-18.286$ \\
2017eaw     & $-1.087(69)$  & $-17.265(203)$  & $-17.828$ \\
2017gmr     & $-0.857(93)$  & $-17.655(235)$  & $-18.005$ \\
2018aoq     & $-2.022(58)$  & $-15.613(136)$  & $-16.029$ \\
2018cuf     & $-1.417(121)$ & $-16.898(359)$  & $-17.224$ \\
2018hwm     & $-2.205(124)$ & $-15.041(284)$  & $-15.191$ 
\\
\hline
\multicolumn{4}{b{0.95\columnwidth}}{\textit{Note}. Numbers in parentheses are $1\,\sigma$ errors in units of 0.001.}
\end{tabular}
\end{table}

Among the SNe used in this work, 44 have nebular spectra (1) between 190 and 410\,d since explosion; (2) being covered by photometry in at least one of these filters: Johnson-Kron-Cousins $V\!RI$ or Sloan $ri$; and (3) with a wavelength coverage enough to compute synthetic magnitudes for the photometric filters mentioned above. For these SNe, their nebular spectra and photometry are useful to estimate $\loi$. The sample of 44~SNe~II is listed in Table~\ref{table:spec_data}, which includes the SN name (Column~1), the heliocentric redshift $z$ (Column~2), $t_\text{expl}$ (Column~3), $\mu$ (Column~4), the Galactic reddening $E_\bv^\text{MW}$ (Column~5), $E_\bv^\text{host}$ (Column~6), the number of selected spectra $N_\text{spec}$ (Column~7), and references for spectroscopic data (Columns~8). Heliocentric redshifts and Galactic reddenings are taken from \citet{2021MNRAS.505.1742R}, while for SN~2015bs I adopt $z=0.027$ \citep{2018NatAs...2..574A} and $E_\bv^\text{MW}=0.044\pm0.007$ \citep{2011ApJ...737..103S}. The photometry I use is the same as that used in \citet{2021MNRAS.505.1742R}, while for SN~2015bs I use the photometry of \citet{2018NatAs...2..574A}.

\begin{table*}
\caption{Sample of SNe with useful nebular spectroscopy.}
\label{table:spec_data}
\begin{tabular}{lccccccc}
\hline
 SN & $cz$ (km\,s$^{-1}$) & $t_\text{expl}$ (MJD) & $\mu$ (mag) &  $E_\bv^\text{MW}$ (mag) & $E_\bv^\text{host}$ (mag) & $N_\text{spec}$ & References$^\dagger$\\
\hline
1990E       & $1362$ & $47934.4\pm1.4$  & $30.83\pm0.26$ & $0.022\pm0.003$ & $ 0.598\pm0.072$ & $ 7$ & 1, 2, 3    \\
1990K       & $1584$ & $48013.8\pm4.2$  & $31.57\pm0.24$ & $0.012\pm0.002$ & $ 0.227\pm0.034$ & $ 2$ & 4, 5       \\
1991G       & $ 757$ & $48281.5\pm5.3$  & $30.76\pm0.17$ & $0.017\pm0.003$ & $ 0.025\pm0.071$ & $ 1$ & 6          \\
1992H       & $1793$ & $48656.4\pm4.5$  & $32.07\pm0.17$ & $0.015\pm0.002$ & $ 0.167\pm0.123$ & $ 6$ & 5, 7, 8    \\
1994N       & $2940$ & $49453.9\pm4.5$  & $33.24\pm0.20$ & $0.032\pm0.005$ & $ 0.045\pm0.036$ & $ 1$ & 9          \\
1996W       & $1617$ & $50180.2\pm2.5$  & $31.86\pm0.19$ & $0.036\pm0.006$ & $ 0.260\pm0.054$ & $ 3$ & 10         \\
1997D       & $1217$ & $50361.0\pm15.0$ & $30.93\pm0.25$ & $0.017\pm0.003$ & $ 0.090\pm0.111$ & $ 1$ & 11         \\
1999em      & $ 800$ & $51474.5\pm2.0$  & $30.31\pm0.09$ & $0.035\pm0.006$ & $ 0.082\pm0.034$ & $ 4$ & 12, 13     \\
1999ga      & $1466$ & $51419.5\pm20.0$ & $31.51\pm0.05$ & $0.173\pm0.028$ & $ 0.511\pm0.084$ & $ 1$ & 14         \\
2002hh      & $ 110$ & $52575.6\pm2.5$  & $29.44\pm0.09$ & $1.065\pm0.046$ & $ 1.545\pm0.182$ & $ 2$ & 15, 16     \\
2003B       & $1141$ & $52622.2\pm4.2$  & $30.62\pm0.25$ & $0.023\pm0.004$ & $ 0.023\pm0.033$ & $ 1$ & 17         \\
2003gd      & $ 657$ & $52716.5\pm21.0$ & $29.95\pm0.08$ & $0.060\pm0.010$ & $ 0.144\pm0.040$ & $ 1$ & 16         \\
2004A       & $ 852$ & $53012.5\pm1.7$  & $30.87\pm0.26$ & $0.013\pm0.002$ & $ 0.177\pm0.043$ & $ 1$ & 5          \\
2004dj      & $ 221$ & $53180.6\pm15.6$ & $27.46\pm0.11$ & $0.034\pm0.006$ & $ 0.094\pm0.035$ & $ 7$ & 5, 18      \\
2004et      & $  40$ & $53270.5\pm0.3$  & $29.44\pm0.09$ & $0.293\pm0.047$ & $ 0.073\pm0.043$ & $20$ & 16, 19, 20 \\
2005ay      & $ 850$ & $53450.7\pm1.8$  & $30.68\pm0.21$ & $0.018\pm0.003$ & $ 0.035\pm0.037$ & $ 1$ & 16         \\
2005cs      & $ 463$ & $53548.4\pm0.3$  & $29.67\pm0.07$ & $0.032\pm0.005$ & $ 0.124\pm0.037$ & $ 6$ & 16, 21     \\
2006my      & $ 788$ & $53942.5\pm20.0$ & $31.40\pm0.10$ & $0.012\pm0.002$ & $-0.062\pm0.152$ & $ 4$ & 5, 20      \\
2006ov      & $1566$ & $53973.5\pm6.0$  & $31.40\pm0.63$ & $0.019\pm0.003$ & $ 0.275\pm0.105$ & $ 1$ & 22         \\
2007it      & $1193$ & $54348.0\pm0.6$  & $30.56\pm0.24$ & $0.099\pm0.016$ & $ 0.109\pm0.111$ & $ 6$ & 17         \\
2008bk      & $ 230$ & $54547.1\pm2.6$  & $27.66\pm0.08$ & $0.017\pm0.003$ & $ 0.090\pm0.111$ & $ 4$ & 17         \\
2008gz      & $1862$ & $54693.5\pm5.0$  & $32.22\pm0.15$ & $0.036\pm0.006$ & $-0.031\pm0.040$ & $ 1$ & 23         \\
2009N       & $ 905$ & $54850.1\pm3.4$  & $30.90\pm0.38$ & $0.018\pm0.003$ & $ 0.265\pm0.045$ & $ 2$ & 24         \\
2009dd      & $ 757$ & $54916.0\pm4.2$  & $30.76\pm0.17$ & $0.017\pm0.003$ & $ 0.285\pm0.057$ & $ 2$ & 10         \\
2009ib      & $1304$ & $55040.3\pm4.0$  & $31.41\pm0.05$ & $0.026\pm0.004$ & $ 0.147\pm0.038$ & $ 2$ & 25         \\
2011fd      & $2101$ & $55777.9\pm4.5$  & $32.41\pm0.22$ & $0.063\pm0.010$ & $ 0.090\pm0.111$ & $ 1$ & 5          \\
2012A       & $ 753$ & $55929.3\pm2.6$  & $30.66\pm0.24$ & $0.027\pm0.004$ & $ 0.040\pm0.049$ & $ 1$ & 5          \\
2012aw      & $ 778$ & $56002.0\pm0.5$  & $29.93\pm0.05$ & $0.024\pm0.004$ & $ 0.115\pm0.041$ & $ 6$ & 5, 26, 27  \\
2012ec      & $1407$ & $56144.5\pm3.4$  & $31.07\pm0.25$ & $0.023\pm0.004$ & $ 0.102\pm0.048$ & $ 2$ & 5, 28      \\
2013K       & $2418$ & $56294.5\pm4.2$  & $32.48\pm0.21$ & $0.122\pm0.019$ & $ 0.512\pm0.066$ & $ 1$ & 29         \\
2013am      & $1114$ & $56371.8\pm0.4$  & $30.36\pm0.29$ & $0.021\pm0.003$ & $ 0.536\pm0.069$ & $ 1$ & 5          \\
2013by      & $1144$ & $56401.6\pm3.6$  & $30.46\pm0.29$ & $0.188\pm0.030$ & $ 0.196\pm0.100$ & $ 1$ & 30         \\
2013ej      & $ 657$ & $56496.8\pm0.2$  & $29.95\pm0.08$ & $0.060\pm0.010$ & $ 0.044\pm0.040$ & $ 3$ & 31, 32     \\
2014G       & $1160$ & $56669.3\pm0.8$  & $31.96\pm0.14$ & $0.010\pm0.002$ & $ 0.268\pm0.046$ & $ 1$ & 33         \\
2014cx      & $1646$ & $56901.9\pm0.3$  & $31.51\pm0.22$ & $0.096\pm0.015$ & $-0.021\pm0.045$ & $ 2$ & 34         \\
ASASSN-14ha & $1504$ & $56909.5\pm0.6$  & $30.86\pm0.15$ & $0.008\pm0.001$ & $ 0.090\pm0.111$ & $ 1$ & 35         \\
2015ba      & $2383$ & $57347.5\pm4.9$  & $32.80\pm0.16$ & $0.015\pm0.002$ & $ 0.416\pm0.047$ & $ 1$ & 36         \\
2015bs      & $8100$ & $56921.5\pm2.6$  & $35.10\pm0.11$ & $0.044\pm0.007$ & $ 0.000\pm0.100$ & $ 1$ & 37         \\
ASASSN-15oz & $2078$ & $57259.1\pm1.9$  & $31.90\pm0.26$ & $0.078\pm0.013$ & $ 0.230\pm0.056$ & $ 3$ & 38         \\
2016aqf     & $1204$ & $57442.6\pm0.3$  & $31.01\pm0.25$ & $0.047\pm0.008$ & $ 0.180\pm0.100$ & $ 5$ & 39         \\
2016gfy     & $2416$ & $57641.3\pm2.6$  & $32.64\pm0.22$ & $0.086\pm0.014$ & $ 0.163\pm0.045$ & $ 2$ & 40         \\
2017eaw     & $  40$ & $57886.2\pm0.6$  & $29.44\pm0.09$ & $0.293\pm0.047$ & $ 0.059\pm0.037$ & $ 4$ & 41         \\
2018cuf     & $3248$ & $58291.8\pm0.3$  & $33.10\pm0.19$ & $0.028\pm0.004$ & $ 0.221\pm0.100$ & $ 1$ & 42         \\
2018hwm     & $2684$ & $58424.8\pm0.9$  & $33.08\pm0.19$ & $0.022\pm0.004$ & $ 0.150\pm0.069$ & $ 1$ & 43         \\
\hline
\multicolumn{8}{l}{\textit{Note}. Quoted uncertainties are $1\,\sigma$ errors.}\\
\multicolumn{8}{b{0.95\textwidth}}{$^\dagger$(1) \citet{1993AJ....105.2236S}; (2) \citet{1994AA...285..147B}; (3) \citet{2000AJ....120..367G}; (4) \citet{1995AA...293..723C}; (5) \citet{2017MNRAS.467..369S}; (6) \citet{1995AJ....110.2868B}; (7) \citet{1996AJ....111.1286C}; (8) \citet{1997ARAA..35..309F}; (9) \citet{2004MNRAS.347...74P}; (10) \citet{2013AA...555A.142I}; (11) \citet{2001MNRAS.322..361B}; (12) \citet{2002PASP..114...35L}; (13) \citet{2003MNRAS.338..939E}; (14) \citet{2009AA...500.1013P}; (15) \citet{2004NewAR..48..595M}; (16) \citet{2014MNRAS.442..844F}; (17) \citet{2017ApJ...850...89G}; (18) \citet{2006Natur.440..505L}; (19) \citet{2006MNRAS.372.1315S}; (20) \citet{2010MNRAS.404..981M}; (21) \citet{2009MNRAS.394.2266P}; (22) \citet{2014MNRAS.439.2873S}; (23) \citet{2011MNRAS.414..167R}; (24) \citet{2014MNRAS.438..368T}; (25) \citet{2015MNRAS.450.3137T}; (26) \citet{2013MNRAS.433.1871B}; (27) \citet{2014MNRAS.439.3694J}; (28) \citet{2015MNRAS.448.2482J}; (29) \citet{2018MNRAS.475.1937T}; (30) \citet{2017ApJ...848....5B}; (31) \citet{2016MNRAS.461.2003Y}; (32) Berkeley SuperNova Database \citep[SNDB;][]{2012MNRAS.425.1789S}; (33) \citet{2016MNRAS.462..137T}; (34) \citet{2016ApJ...832..139H}; (35) Public ESO Spectroscopic Survey for Transient Objects Survey \citep[PESSTO;][]{2015AA...579A..40S}; (36) \citet{2018MNRAS.479.2421D}; (37) \citet{2018NatAs...2..574A}; (38) \citet{2019MNRAS.485.5120B}; (39) \citet{2020MNRAS.497..361M}; (40) \citet{2019ApJ...882...68S}; (41) \citet{2019ApJ...875..136V}; (42) \citet{2020ApJ...906...56D}; (43) \citet{2021MNRAS.501.1059R}.}
\end{tabular}
\end{table*}

As in \citet{2021MNRAS.505.1742R}, for our Galaxy and host galaxies I assume the extinction curve $R_\lambda=A_\lambda/E_\bv$ of \citet{1999PASP..111...63F} with ${R_V=A_V/E_\bv}$ of $3.1$, while for SN~2002hh I adopt a host galaxy $R_V$ of $1.1$.

\subsection{Progenitor sample}
Twelve SNe in my sample have photometry of their confirmed or candidate progenitors.\footnote{I do not include SN~2009md because the source identified as its progenitor by \citet{2011MNRAS.417.1417F} is still present in images taken three years after the SN explosion \citep{2015MNRAS.447.3207M}.} Difference images between pre-explosion and late-time images, and the corresponding progenitor photometry (in Vega magnitudes) are available for SNe~2003gd, 2004et, 2004A, 2005cs, 2006my, 2008bk, and 2012aw. The progenitor candidates for SNe~2012A, 2012ec, 2013ej, 2017eaw, and 2018aoq are still not confirmed by their disappearance in late-time images, so the reported progenitor photometry is not definitive. For completeness, I include six SNe~II for which $3\,\sigma$ detection limits for their progenitors are available: SNe~1999em, 2002hh ($5\,\sigma$ detection limit), 2006ov, 2007aa, 2009hd,\footnote{As in \citet{2015PASA...32...16S}, I adopt the progenitor magnitude of SN~2009hd as an upper limit because it is close to the $3\,\sigma$ detection limit \citep[see][]{2011ApJ...742....6E}.} and 2009ib.\footnote{I assume the scenario where the progenitor is not the yellow source detected at the SN position but a RSG too faint to be detected \citep[see][]{2015MNRAS.450.3137T}.} The list of the 18~SNe~II with progenitor photometry or upper limits is summarized in Table~\ref{table:m_progenitors}. This includes the SN name (Column~1), $\mu$ (Column~2), $E_\bv^\text{MW}$ (Column~3), $E_\bv^\text{host}$ (Column~4), the filter used to observe the progenitor ($x$), the apparent progenitor magnitude ($m_{\text{prog},x}$), and its reference (Columns~5, 6, and 7, respectively).

\begin{table*}
\caption{SN~II progenitor sample}
\label{table:m_progenitors}
\begin{tabular}{lcccccc}
\hline
 SN      & $\mu$       & $E_\bv^\text{MW}$& $E_\bv^\text{host}$ & $x$     & $m_{\text{prog},x}$    & Reference   \\
\hline
 1999em  & $30.31\pm0.09$ & $0.035\pm0.006$ & $ 0.082\pm0.034$ & $I$              & >$ 23.0$       & \citet{2009MNRAS.395.1409S} \\
 2002hh  & $29.44\pm0.09$ & $1.065\pm0.046$ & $ 1.545\pm0.182$ & $i'_\text{Gunn}$ & >$ 22.8$       & \citet{2009MNRAS.395.1409S} \\
 2003gd  & $29.95\pm0.08$ & $0.060\pm0.010$ & $ 0.144\pm0.040$ & $I_\text{J}$     & $23.14\pm0.08$ & \citet{2009Sci...324..486M} \\
 2004A   & $30.87\pm0.26$ & $0.013\pm0.002$ & $ 0.177\pm0.043$ & WFPC2 F814W      & $24.36\pm0.12$ & \citet{2014MNRAS.438..938M} \\
 2004et  & $29.44\pm0.09$ & $0.293\pm0.047$ & $ 0.073\pm0.043$ & $I_\text{J}$     & $21.88\pm0.17$ & \citet{2011MNRAS.410.2767C} \\
 2005cs  & $29.67\pm0.07$ & $0.032\pm0.005$ & $ 0.124\pm0.037$ & ACS/WFC F814W    & $23.62\pm0.07$ & \citet{2014MNRAS.438..938M} \\
 2006my  & $31.40\pm0.10$ & $0.012\pm0.002$ & $-0.062\pm0.152$ & WFPC2 F814W      & $24.86\pm0.13$ & \citet{2014MNRAS.438..938M} \\
 2006ov  & $31.40\pm0.63$ & $0.019\pm0.003$ & $ 0.275\pm0.105$ & WFPC2 F814W      & >$ 24.2$       & \citet{2011MNRAS.410.2767C} \\
 2007aa  & $31.99\pm0.27$ & $0.023\pm0.004$ & $ 0.046\pm0.100$ & WFPC2 F814W      & >$24.44$       & \citet{2009MNRAS.395.1409S} \\
 2008bk  & $27.66\pm0.08$ & $0.017\pm0.003$ & $ 0.090\pm0.111$ & $K_\text{S}$     & $18.39\pm0.03$ & \citet{2014MNRAS.438.1577M} \\
 2009hd  & $30.15\pm0.07$ & $0.029\pm0.005$ & $ 1.206\pm0.056$ & WFPC2 F814W      & >$23.54$       & \citet{2011ApJ...742....6E} \\
 2009ib  & $31.41\pm0.05$ & $0.026\pm0.004$ & $ 0.147\pm0.038$ & WFPC2 F814W      & >$23.25$       & \citet{2015MNRAS.450.3137T} \\
 2012A   & $30.66\pm0.24$ & $0.027\pm0.004$ & $ 0.040\pm0.049$ & $K'$             & $20.29\pm0.13$ & \citet{2013MNRAS.434.1636T} \\
 2012aw  & $29.93\pm0.05$ & $0.024\pm0.004$ & $ 0.115\pm0.041$ & $K_\text{S}$     & $19.56\pm0.29$ & \citet{2016MNRAS.456L..16F} \\
 2012ec  & $31.07\pm0.25$ & $0.023\pm0.004$ & $ 0.102\pm0.048$ & WFPC2 F814W      & $23.39\pm0.18$ & \citet{2013MNRAS.431L.102M} \\
 2013ej  & $29.95\pm0.08$ & $0.060\pm0.010$ & $ 0.044\pm0.040$ & ACS/WFC F814W    & $22.66\pm0.03$ & \citet{2014MNRAS.439L..56F} \\
 2017eaw & $29.44\pm0.09$ & $0.293\pm0.047$ & $ 0.059\pm0.037$ & WFC3/IR F160W    & $19.36\pm0.01$ & \citet{2019ApJ...875..136V} \\
 2018aoq & $30.99\pm0.06$ & $0.023\pm0.004$ & $ 0.086\pm0.040$ & WFC3/IR F160W    & $21.84\pm0.05$ & \citet{2019AA...622L...1O} \\
\hline
\multicolumn{7}{l}{\textit{Note}. Quoted uncertainties are $1\,\sigma$ errors.}
\end{tabular}
\end{table*}

\subsection{RSG samples}
For the comparison between luminosities of SN~II progenitors and RSGs, I use the RSG samples reported by \citet{2020ApJ...900..118N} for LMC, by \citet{2021ApJ...922..177M} for SMC, and by \citet{2021AJ....161...79M} for M31 and M33. The authors used $JK_\text{S}$ photometry to identify RSGs in the colour-magnitude diagram and to compute luminosities using the BC technique. The RSG samples of LMC, M31, and M33 (SMC) are complete to a luminosity limit of $\log(L/\lsun)=4.0$\,dex (3.7\,dex). For this work, I re-compute RSG luminosities using the most recent and precise distances reported for the galaxies mentioned above. For LMC and SMC I adopt distance moduli of $18.477\pm0.024$ \citep{2019Natur.567..200P} and $18.977\pm0.032$\,mag \citep{2020ApJ...904...13G}, respectively, which are based on late-type eclipsing binary stars. For M31 and M33 I adopt distance moduli of $24.407\pm0.032$ \citep{2021ApJ...920...84L} and $24.568\pm0.064$\,mag \citep{2021ApJ...916...19Z}, respectively, estimated using the near-infrared Cepheid period-luminosity relation and the $J$-region Asymptotic Giant Branch method, respectively. As \citet{2021AJ....161...79M} mentioned, the M31 sample includes stars with unlikely large extinction values and therefore unrealistic high luminosities. In order not to include those stars in the analysis, I remove stars with $A_V\geq2.5$\,mag (around 0.4~per~cent of the total sample).

\section{Methodology}\label{sec:methodology}

\subsection
[{[O I] line luminosity}]
{[O\,{\sevensize I}] $\bmath{\lambda\lambda}$6300, 6364 line luminosity}

To compute $L_{[\ion{O}{I}]}$, it is necessary to calibrate the flux of the spectra in the wavelength region of the [\ion{O}{I}] doublet. For this, I scale each spectrum by a constant $C$ such that $\log C=0.4\langle m_{x,\text{syn}}-m_{x,t_\text{spec}} \rangle$. Here, $m_{x,\text{syn}}$ is the $x$-band synthetic magnitude computed from the spectrum, $m_{x,t_\text{spec}}$ is the SN apparent magnitude at the epoch of the spectrum ($t_\text{spec}$), and angle brackets denote an average over the bands used to estimate $C$. To compute synthetic magnitudes I use the methodology of \citet{2021MNRAS.505.1742R}, while to interpolate photometry to the epochs of the spectra I use the \texttt{ALR} code\footnote{\url{https://github.com/olrodrig/ALR}} \citep{2019MNRAS.483.5459R}. This code performs \texttt{loess} non-parametric regressions \citep{Cleveland_etal1992} to the input data, taking into account observed and intrinsic errors, along with the presence of possible outliers. For the flux calibration of the spectra, I use $V$ and $r/R$ photometry. If it is not possible to calculate $m_{x,\text{syn}}-m_{x,t}$ for one of those bands, then the $i/I$ band is included. In the case of SN~2015bs, I extrapolate its $R$-band photometry (see Appendix~\ref{sec:SNe_appendix}) to the epoch of the spectrum using a straight line fit.

Once the flux of each spectrum is calibrated, the flux of the [\ion{O}{I}]\,$\lambda\lambda$6300,~6364 line is given by
\begin{equation}
F_{[\ion{O}{I}]} = h\sum_{i=1}^{n}(f_i-f_{c,\lambda_i}).
\end{equation}
Here, $f_i$ and $f_{c,\lambda_i}$ are the observed and continuum flux at wavelength $\lambda_i$, respectively, $h$ is the spectral dispersion, and $n$ is the number of pixels between the blue and red endpoints of the [\ion{O}{I}] doublet ($\lambda_1$ and $\lambda_n$, respectively).

\begin{figure}
\includegraphics[width=1.0\columnwidth]{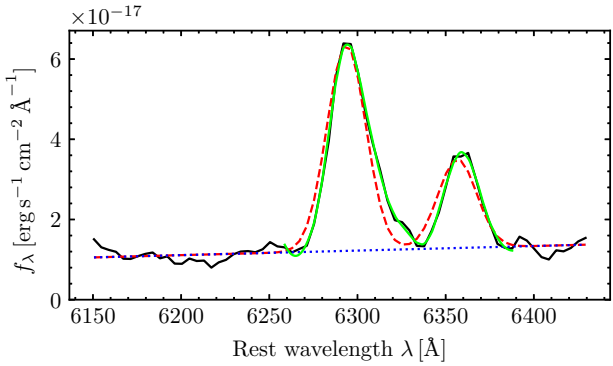}
\caption{Spectrum of SN~2009N at 370\,d since explosion showing the [\ion{O}{I}]\,$\lambda\lambda$6300, 6364 line. The red dashed line is the double Gaussian fit plus a straight line (blue dotted line), while the green solid line corresponds to the \texttt{ALR} fit.}
\label{fig:FOI}
\end{figure}

To estimate the flux of the continuum, I fit the [\ion{O}{I}] doublet with a double Gaussian separated by 6.4\,nm plus a straight line corresponding to $f_{c,\lambda}$ \citep[e.g.][]{2011AcA....61..179E}. Fig.~\ref{fig:FOI} shows this analytical fit applied to the [\ion{O}{I}] doublet of SN~2009N at 370\,d since the explosion. I adopt as $\lambda_1$ and $\lambda_n$ the wavelengths for which the extremes of the double Gaussian are equal to one~per~cent of the maximum.

The error on $\log F_{[\ion{O}{I}]}$ is given by
\begin{equation}
\sigma_{\log F_{[\ion{O}{I}]}}=\sqrt{\left[\frac{h}{\ln(10)F_{[\ion{O}{I}]}}\right]^2\sum_{i=1}^{n}(\sigma_{f_i}^2+\sigma_{f_{c,\lambda_i}}^2)  +\frac{\sigma_m^2}{6.25}},
\end{equation}
where $\sigma_{f_i}$ and $\sigma_{f_{c,\lambda_i}}$ are the errors on $f_i$ and $f_{c,\lambda_i}$, respectively, and $\sigma_m$ is the mean error of the photometry used to calibrate the spectrum. To estimate $\sigma_{f_i}$, I fit the [\ion{O}{I}] doublet with the \texttt{ALR} code, and then assume the sample standard deviation ($\ssd$) around the \texttt{ALR} fit as the error on $\sigma_{f_i}$. I also assume $\sigma_{f_{c,\lambda_i}}=\sigma_{f_i}$.

\begin{figure}
\includegraphics[width=1.0\columnwidth]{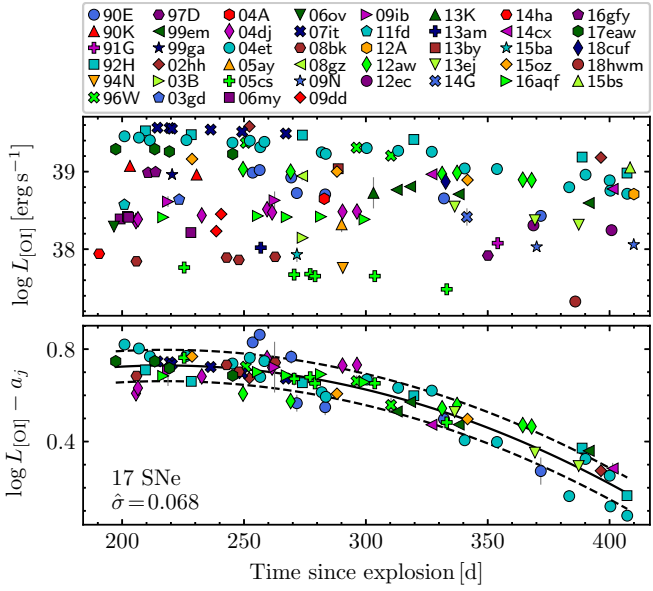}
\caption{Top panel: [\ion{O}{I}]\,$\lambda\lambda$6300, 6364 line luminosity as a function of time since explosion. Bottom panel: $\log L_{[\ion{O}{I}]}-a_j$ against time since explosion for the 17~SNe used to estimate $\Psi(t)$. The solid line is a quadratic fit corresponding to $\Psi(t)$, while dashed lines are $\pm1\,\ssd$ limits.}
\label{fig:lLOI_vs_phase}
\end{figure}

The luminosity of the [\ion{O}{I}] doublet (in units of erg\,s$^{-1}$) is given by
\begin{equation}
\log L_{[\ion{O}{I}]} = \log F_{[\ion{O}{I}]} + (\mu+R_{6332}E_\bv)/2.5 + 40.078,
\end{equation}
where $F_{[\ion{O}{I}]}$ is in units of erg\,s$^{-1}$\,cm$^{-2}$, $E_\bv$ is the total reddening, $R_{6332}=2.49$ is the $R_\lambda$ value at $\lambda=6332$\,\AA\ (the middle wavelength of the [\ion{O}{I}] doublet),\footnote{In the case of SN~2002hh, the term $R_{6332}E_\bv$ is replaced by $2.49E_\bv^\text{MW}+R_{6332}E_\bv^\text{host}$, where $R_{6332}=0.53$ for $R_V=1.1$.} and the constant term provides the conversion from magnitude to cgs units. The $\log L_{[\ion{O}{I}]}$ values are shown in the top panel of Fig.~\ref{fig:lLOI_vs_phase} against time since explosion $t=(t_\text{spec}-t_\text{expl})/(1+z)$.

To calculate $\log\loi$, I construct an analytical expression for $\log L_{[\ion{O}{I}]}$ as a function of $t$, $\log L_{[\ion{O}{I}]}(t)$, which is then evaluated at $t=350$\,d. For each SN, I model $\log L_{[\ion{O}{I}]}(t)$ as
\begin{equation}\label{eq:LOI_t}
\log L_{[\ion{O}{I}]}(t) = \delta + \Psi(t),
\end{equation}
where $\delta$ is the vertical intercept of $\log L_{[\ion{O}{I}]}(t)$, and $\Psi(t)$ is a polynomial representing the dependence of $\log L_{[\ion{O}{I}]}(t)$ on $t$ (i.e. the shape of the curve). Under the assumption that the $\log L_{[\ion{O}{I}]}(t)$ curves of all SNe~II have the same shape, the parameters of $\Psi(t)$ can be computed minimizing
\begin{equation}\label{eq:s2}
s^2=\sum_{j=1}^{N_\text{SNe}}\sum_{i=1}^{N_{\text{spec},j}} [\log L_{{[\ion{O}{I}]}_{j,i}} -a_j-\Psi(t_{j,i})]^2.
\end{equation}
Here, $N_\text{SNe}$ is the number of SNe, $a_j$ is an additive term to normalize the $\log L_{[\ion{O}{I}]}$ values of each SN to the same scale, while the polynomial order of $\Psi(t)$ is determined with the Bayesian information criterion \citep{1978AnSta...6..461S}. To estimate $\Psi(t)$, I use 17~SNe having two or more $\log L_{[\ion{O}{I}]}$ measurements covering a time range of at least 40\,d.

The bottom panel of Fig.~\ref{fig:lLOI_vs_phase} shows the result of the minimization of equation~(\ref{eq:s2}). I find that $\Psi(t)$ is quadratic on $t$, given by
\begin{equation}\label{eq:psi}
\Psi(t) = 0.670\left(\frac{t}{100\,\text{d}}\right) -0.154\left(\frac{t}{100\,\text{d}}\right)^2,
\end{equation}
with $\ssd=0.068$\,dex. Once the shape of $\log L_{[\ion{O}{I}]}(t)$ is known, the $\delta$ value for each SN can be computed using the weighted mean
\begin{equation}
\delta = \sum_{i=1}^{N_\text{spec}} (w_i\cdot[\log L_{{[\ion{O}{I}]}_i}-\Psi(t_i)])/\sum_{i=1}^{N_\text{spec}} w_i,
\end{equation}
with weights $w_i = 1/(\sigma_{\log F_{{[\ion{O}{I}]}_i}}^2 + \ssd^2)$. The error on $\delta$ is given by the weighted mean error.

I compute $\log \loi$ using equations~(\ref{eq:LOI_t}) and (\ref{eq:psi}), and the corresponding $\delta$ value. Those estimates are listed in Table~\ref{table:lLOI_350d} along with their errors, given by
\begin{equation}
\sigma_{\log\loi} = \sqrt{\sigma_\delta^2 + \frac{\sigma_\mu^2+R_{6332}^2\sigma_{E_\bv}^2}{6.25}+0.41^2\left[\frac{\sigma_{t_\text{expl}}}{100\,\text{d}}\right]^2}.
\end{equation}
The mean $\log\loi$ error is of 0.13\,dex, of which 65~per~cent is induced by errors on distance and host galaxy reddening.

\begin{table}
\caption{[\ion{O}{I}]\,$\lambda\lambda$6300, 6364 line luminosities at 350\,d.}
\label{table:lLOI_350d}
\begin{tabular}{lclc}
\hline
 SN         & $\log\loi$       & SN          & $\log\loi$ \\
\hline
1990E       & $38.627\pm0.130$ & 2009N       & $38.241\pm0.167$ \\
1990K       & $38.751\pm0.114$ & 2009dd      & $38.085\pm0.105$ \\
1991G       & $38.095\pm0.137$ & 2009ib      & $38.359\pm0.088$ \\
1992H       & $39.276\pm0.144$ & 2011fd      & $38.308\pm0.161$ \\
1994N       & $37.571\pm0.125$ & 2012A       & $39.010\pm0.131$ \\
1996W       & $39.106\pm0.103$ & 2012aw      & $38.885\pm0.053$ \\
1997D       & $37.919\pm0.184$ & 2012ec      & $38.440\pm0.125$ \\
1999em      & $38.695\pm0.061$ & 2013K       & $38.569\pm0.233$ \\
1999ga      & $38.692\pm0.141$ & 2013am      & $37.773\pm0.153$ \\
2002hh      & $38.926\pm0.086$ & 2013by      & $38.845\pm0.181$ \\
2003B       & $37.927\pm0.128$ & 2013ej      & $38.481\pm0.065$ \\
2003gd      & $38.370\pm0.121$ & 2014G       & $38.382\pm0.144$ \\
2004A       & $38.446\pm0.132$ & 2014cx      & $38.950\pm0.112$ \\
2004dj      & $38.213\pm0.089$ & ASASSN-14ha & $37.682\pm0.152$ \\
2004et      & $39.095\pm0.074$ & 2015ba      & $37.705\pm0.138$ \\
2005ay      & $38.131\pm0.147$ & 2015bs      & $39.344\pm0.149$ \\
2005cs      & $37.460\pm0.055$ & ASASSN-15oz & $38.853\pm0.126$ \\
2006my      & $38.089\pm0.180$ & 2016aqf     & $38.187\pm0.145$ \\
2006ov      & $38.027\pm0.284$ & 2016gfy     & $38.720\pm0.112$ \\
2007it      & $39.277\pm0.150$ & 2017eaw     & $39.000\pm0.078$ \\
2008bk      & $37.619\pm0.121$ & 2018cuf     & $38.805\pm0.146$ \\
2008gz      & $38.722\pm0.113$ & 2018hwm     & $37.492\pm0.140$ \\
\hline
\multicolumn{4}{b{0.95\columnwidth}}{\textit{Notes}. $\loi$ values are in units of erg\,s$^{-1}$. Quoted uncertainties are $1\,\sigma$ errors.}
\end{tabular}
\end{table}

\subsection{Progenitor luminosity from pre-explosion photometry}\label{sec:Lf_images}
For the SNe listed in Table~\ref{table:m_progenitors}, I compute $\Lf$ (in units of $\lsun$) using the BC technique
\begin{equation}\label{eq:Lf_BC}
\log \Lf = (\mu-\text{BC}_x-m_{\text{prog},x}+R_{\lambda_x} E_\bv+4.74)/2.5,
\end{equation}
where BC$_x$ and $\lambda_x$ are the $x$-band BC and effective wavelength, respectively.\footnote{In the case of SN~2002hh, the term $R_{\lambda_x}E_\bv$ is replaced by $R_{\lambda_x}E_\bv^\text{MW}+R_{\lambda_x}^\prime E_\bv^\text{host}$, where $R_{\lambda_x}^\prime$ is the $R_{\lambda_x}$ value for $R_V=1.1$.} In this work I use the empirical BCs for RSGs presented in \citet{2018MNRAS.474.2116D,2020MNRAS.493..468D}. For 16 of the 18 SN~II progenitors listed in Table~\ref{table:m_progenitors}, I adopt the BC$_x$ estimates reported by \citet{2018MNRAS.474.2116D,2020MNRAS.493..468D}. For the progenitor of SN~2009ib, I adopt the average $\mathrm{BC_{F814W}}$ value for late-type RSGs of $0.00\pm0.15$ computed by \citet{2018MNRAS.474.2116D}. For the progenitor of SN~2003gd, identified as a M0 to M2 RSG \citep{2009Sci...324..486M}, I compute a Johnson $I$-band BC of $0.60\pm0.17$ by averaging the $\text{BC}_I$ values for M0-M2 RSGs presented in Fig.~2 of \citet{2018MNRAS.474.2116D}. The adopted BC$_x$ values are listed in Table~\ref{table:BC}.

\begin{table}
\caption{Adopted bolometric corrections.}
\label{table:BC}
\begin{tabular}{lclc}
\hline
 SN      & BC$_x$         & SN      & BC$_x$        \\
\hline
 1999em  & $-0.32\pm0.15$ & 2008bk  & $3.00\pm0.18$ \\
 2002hh  & $-0.49\pm0.15$ & 2009hd  & $0.00\pm0.15$ \\
 2003gd  & $ 0.60\pm0.17$ & 2009ib  & $0.00\pm0.15$ \\
 2004A   & $ 0.00\pm0.15$ & 2012A   & $3.00\pm0.18$ \\
 2004et  & $ 0.25\pm0.15$ & 2012aw  & $3.00\pm0.18$ \\
 2005cs  & $ 0.05\pm0.15$ & 2012ec  & $0.00\pm0.15$ \\
 2006my  & $ 0.00\pm0.15$ & 2013ej  & $0.40\pm0.20$ \\
 2006ov  & $ 0.00\pm0.15$ & 2017eaw & $2.60\pm0.10$ \\
 2007aa  & $ 0.00\pm0.15$ & 2018aoq & $2.60\pm0.10$ \\
\hline
\multicolumn{4}{l}{\textit{Note}. Quoted uncertainties are $1\,\sigma$ errors.}
\end{tabular}
\end{table}

The $x$-band effective wavelength is defined as
\begin{equation}
\lambda_x=\frac{\int d\lambda S_{x,\lambda} \lambda^2 f_{\lambda}}{\int d\lambda S_{x,\lambda} \lambda f_{\lambda}}
\end{equation}
\citep[e.g.][]{2012PASP..124..140B}, where $S_{x,\lambda}$ is the photon-counting response function for the $x$-band, and $f_{\lambda}$ is the SED of the progenitor. To estimate $\lambda_x$ for the filters listed in Column~5 of Table~\ref{table:m_progenitors}, I use the response functions available in the SVO Filter Profile Service\footnote{\url{http://svo2.cab.inta-csic.es/theory/fps/}} \citep{2012ivoa.rept.1015R,2020sea..confE.182R}, and assume a Planck function as $f_\lambda$, using temperatures between 3400 and 4500\,K to represent the SEDs of RSGs.
Then, I compute $\lambda_x$ estimates for $10^4$ temperature values randomly selected from a uniform distribution between 3400 and 4500\,K, and adopt the mean of these estimates as the final effective wavelength. Table~\ref{table:R_eff} lists the final $\lambda_x$ values and the corresponding $R_{\lambda_x}$ estimates for different bands.

\begin{table}
\caption{Effective wavelengths and $R_{\lambda_x}$ values.}
\label{table:R_eff}
\begin{tabular}{ccc}
\hline
 $x$              & $\lambda_x$ (\AA) & $R_{\lambda_x}$ \\
\hline
 $i'_\text{Gunn}$ &  $7720$ & $1.82^\dagger$ \\
 WFPC2 F814W      &  $8010$ & $1.71$ \\
 $I$              &  $8030$ & $1.70$ \\ 
 ACS/WFC F814W    &  $8050$ & $1.69$ \\
 $I_\text{J}$     &  $8640$ & $1.50$ \\ 
 WFC3/IR F160W    & $15290$ & $0.58$ \\
 $K'$             & $21060$ & $0.36$ \\
 $K_\text{S}$     & $21520$ & $0.35$ \\
\hline
\multicolumn{3}{l}{$^\dagger R_{\lambda_x}=0.31$ for $R_V=1.1$.}
\end{tabular}
\end{table}

The $\log\Lf$ values computed from progenitor photometry ($\log\Lf(\text{phot})$), are listed in Column~2 of Table~\ref{table:Lf}. The mean $\log\Lf(\text{phot})$ error is of 0.10\,dex, of which 46 and 26~per~cent is induced by errors on BC$_x$ and $\mu$, respectively.

\begin{table*}
\caption{Progenitor luminosities.}
\label{table:Lf}
\begin{tabular}{lccccc}
\hline
 SN &$\log\Lf(\text{phot})$ & $\log \Lf(\loi)$  & $\log\Lf(\mni)$ & $\log \Lf(\mv)$ & $\log\Lf$ \\ 
\hline
1980K                   & --               & --              & $4.787\pm0.099$ & $4.984\pm0.109$ & $ 4.787\pm 0.099$ \\
1986I$^{\dagger}$       & --               & --              & $4.867\pm0.119$ & $4.879\pm0.154$ & $ 4.867\pm 0.119$ \\
1988A$^{\dagger}$       & --               & --              & $4.964\pm0.114$ & $4.777\pm0.132$ & $ 4.964\pm 0.114$ \\
1990E                   & --               & $4.792\pm0.072$ & $4.827\pm0.090$ & $4.828\pm0.101$ & $ 4.792\pm 0.072$ \\
1990K                   & --               & $4.839\pm0.068$ & $4.783\pm0.089$ & $4.829\pm0.089$ & $ 4.839\pm 0.068$ \\
1991G$^{\dagger}$       & --               & $4.590\pm0.074$ & $4.628\pm0.083$ & $4.508\pm0.090$ & $ 4.590\pm 0.074$ \\
1991al                  & --               & --              & $4.703\pm0.074$ & $4.654\pm0.071$ & $ 4.654\pm 0.071$ \\
1992H                   & --               & $5.038\pm0.076$ & $5.125\pm0.111$ & $5.061\pm0.116$ & $ 5.038\pm 0.076$ \\
1992ba$^{\dagger}$      & --               & --              & $4.646\pm0.091$ & $4.596\pm0.095$ & $ 4.646\pm 0.091$ \\
1994N                   & --               & $4.391\pm0.071$ & $4.375\pm0.085$ & $4.409\pm0.082$ & $ 4.391\pm 0.071$ \\
1995ad                  & --               & --              & $4.904\pm0.088$ & $4.921\pm0.111$ & $ 4.904\pm 0.088$ \\
1996W                   & --               & $4.973\pm0.066$ & $5.043\pm0.082$ & $5.013\pm0.086$ & $ 4.973\pm 0.066$ \\
1997D$^{\dagger}$       & --               & $4.523\pm0.088$ & $4.485\pm0.098$ & --              & $ 4.523\pm 0.088$ \\
1999ca                  & --               & --              & $4.594\pm0.075$ & $4.853\pm0.072$ & $ 4.853\pm 0.072$ \\
1999em                  & $<5.027\pm0.074$ & $4.817\pm0.058$ & $4.870\pm0.073$ & $4.826\pm0.069$ & $ 4.817\pm 0.058$ \\
1999ga                  & --               & $4.816\pm0.075$ & $4.795\pm0.089$ & --              & $ 4.816\pm 0.075$ \\
1999gi$^{\dagger}$      & --               & --              & $4.852\pm0.079$ & $4.720\pm0.078$ & $ 4.720\pm 0.078$ \\
2001X                   & --               & --              & $4.821\pm0.082$ & $4.744\pm0.082$ & $ 4.744\pm 0.082$ \\
2001dc                  & --               & --              & $4.457\pm0.086$ & $4.434\pm0.091$ & $ 4.457\pm 0.086$ \\
2002gw                  & --               & --              & $4.702\pm0.084$ & $4.664\pm0.084$ & $ 4.702\pm 0.084$ \\
2002hh                  & $<5.715\pm0.081$ & $4.905\pm0.062$ & $4.978\pm0.075$ & $4.872\pm0.091$ & $ 4.905\pm 0.062$ \\
2002hx                  & --               & --              & $4.926\pm0.073$ & $4.821\pm0.072$ & $ 4.821\pm 0.072$ \\
2003B$^{\dagger}$       & --               & $4.526\pm0.072$ & $4.405\pm0.088$ & $4.364\pm0.088$ & $ 4.526\pm 0.072$ \\
2003T                   & --               & --              & $4.846\pm0.083$ & $4.822\pm0.072$ & $ 4.822\pm 0.072$ \\
2003Z                   & --               & --              & $4.385\pm0.088$ & $4.326\pm0.090$ & $ 4.385\pm 0.088$ \\
2003fb                  & --               & --              & $4.777\pm0.086$ & $4.664\pm0.078$ & $ 4.664\pm 0.078$ \\
2003gd                  & $ 4.502\pm0.085$ & $4.694\pm0.070$ & $4.671\pm0.083$ & --              & $ 4.616\pm 0.054$ \\
2003hd                  & --               & --              & $4.837\pm0.073$ & $4.913\pm0.071$ & $ 4.913\pm 0.071$ \\
2003hk                  & --               & --              & $4.732\pm0.084$ & $4.924\pm0.079$ & $ 4.924\pm 0.079$ \\
2003hn                  & --               & --              & $4.812\pm0.072$ & $4.857\pm0.068$ & $ 4.857\pm 0.068$ \\
2003ho                  & --               & --              & $4.717\pm0.084$ & $4.793\pm0.088$ & $ 4.717\pm 0.084$ \\
2003iq                  & --               & --              & $4.859\pm0.078$ & $4.843\pm0.077$ & $ 4.843\pm 0.077$ \\
2004A$^{\dagger}$       & $ 4.630\pm0.133$ & $4.723\pm0.073$ & $4.716\pm0.088$ & $4.636\pm0.092$ & $ 4.702\pm 0.064$ \\
2004dj$^{\dagger}$      & --               & $4.635\pm0.063$ & $4.566\pm0.093$ & $4.623\pm0.074$ & $ 4.635\pm 0.063$ \\
2004eg                  & --               & --              & $4.454\pm0.115$ & --              & $ 4.454\pm 0.115$ \\
2004ej                  & --               & --              & $4.621\pm0.106$ & $4.779\pm0.109$ & $ 4.621\pm 0.106$ \\
2004et                  & $ 5.039\pm0.105$ & $4.969\pm0.060$ & $5.000\pm0.075$ & $5.055\pm0.079$ & $ 4.986\pm 0.052$ \\
2004fx                  & --               & --              & $4.616\pm0.109$ & $4.593\pm0.113$ & $ 4.616\pm 0.109$ \\
2005af                  & --               & --              & $4.780\pm0.111$ & --              & $ 4.780\pm 0.111$ \\
2005au                  & --               & --              & $4.921\pm0.079$ & $4.970\pm0.089$ & $ 4.921\pm 0.079$ \\
2005ay$^{\dagger}$      & --               & $4.604\pm0.077$ & $4.628\pm0.086$ & $4.543\pm0.083$ & $ 4.604\pm 0.077$ \\
2005cs$^{\dagger}$      & $ 4.401\pm0.076$ & $4.349\pm0.057$ & $4.396\pm0.076$ & $4.509\pm0.068$ & $ 4.368\pm 0.046$ \\
2005dx                  & --               & --              & $4.485\pm0.118$ & $4.560\pm0.122$ & $ 4.485\pm 0.118$ \\
2006my                  & $ 4.478\pm0.137$ & $4.588\pm0.086$ & $4.681\pm0.097$ & --              & $ 4.557\pm 0.073$ \\
2006ov                  & $<4.977\pm0.269$ & $4.564\pm0.120$ & $4.491\pm0.149$ & --              & $ 4.564\pm 0.120$ \\
2007aa                  & $<4.963\pm0.141$ & --              & $4.757\pm0.098$ & $4.793\pm0.115$ & $ 4.757\pm 0.098$ \\
2007hv                  & --               & --              & $4.855\pm0.091$ & $4.828\pm0.113$ & $ 4.855\pm 0.091$ \\
2007it                  & --               & $5.038\pm0.078$ & $5.026\pm0.093$ & $4.996\pm0.117$ & $ 5.038\pm 0.078$ \\
2007od                  & --               & --              & --              & $5.005\pm0.100$ & $ 5.005\pm 0.100$ \\
2008K                   & --               & --              & $4.712\pm0.112$ & $4.853\pm0.109$ & $ 4.853\pm 0.109$ \\
2008M                   & --               & --              & $4.715\pm0.109$ & $4.781\pm0.112$ & $ 4.715\pm 0.109$ \\
2008aw                  & --               & --              & $4.982\pm0.087$ & $4.955\pm0.082$ & $ 4.955\pm 0.082$ \\
2008bk$^{\dagger}$      & $ 4.419\pm0.081$ & $4.410\pm0.070$ & $4.485\pm0.081$ & $4.418\pm0.106$ & $ 4.414\pm 0.053$ \\
2008gz                  & --               & $4.828\pm0.068$ & $4.892\pm0.078$ & --              & $ 4.828\pm 0.068$ \\
2008in                  & --               & --              & $4.698\pm0.145$ & $4.692\pm0.166$ & $ 4.698\pm 0.145$ \\
2009N$^{\dagger}$       & --               & $4.645\pm0.083$ & $4.572\pm0.104$ & $4.506\pm0.114$ & $ 4.645\pm 0.083$ \\
2009at                  & --               & --              & $4.643\pm0.088$ & $4.697\pm0.100$ & $ 4.643\pm 0.088$ \\
2009ay                  & --               & --              & $5.053\pm0.088$ & $5.073\pm0.095$ & $ 5.053\pm 0.088$ \\
2009bw                  & --               & --              & $4.649\pm0.089$ & $4.754\pm0.097$ & $ 4.649\pm 0.089$ \\
2009dd                  & --               & $4.586\pm0.066$ & $4.781\pm0.087$ & $4.752\pm0.086$ & $ 4.586\pm 0.066$ \\
2009hd                  & $<5.385\pm0.095$ & --              & $4.538\pm0.074$ & $4.812\pm0.075$ & $ 4.538\pm 0.074$ \\
2009ib                  & $<5.278\pm0.068$ & $4.690\pm0.063$ & $4.840\pm0.076$ & $4.610\pm0.068$ & $ 4.690\pm 0.063$ \\
2009md                  & --               & --              & $4.468\pm0.077$ & $4.508\pm0.076$ & $ 4.508\pm 0.076$ \\
2010aj                  & --               & --              & $4.473\pm0.076$ & $4.843\pm0.077$ & $ 4.473\pm 0.076$ \\

\end{tabular}
\end{table*}
\begin{table*}
\contcaption{}
\begin{tabular}{lccccc}
\hline
 SN &$\log \Lf(\text{phot})$ & $\log \Lf(\loi)$  & $\log\Lf(\mni)$ & $\log \Lf(\mv)$ & $\log\Lf$ \\
\hline
PTF10gva                & --               & --              & $4.963\pm0.096$ & --              & $ 4.963\pm 0.096$ \\
2011fd                  & --               & $4.671\pm0.081$ & $4.769\pm0.091$ & --              & $ 4.671\pm 0.081$ \\
PTF11go                 & --               & --              & $4.721\pm0.104$ & --              & $ 4.721\pm 0.104$ \\
PTF11htj                & --               & --              & $4.864\pm0.110$ & --              & $ 4.864\pm 0.110$ \\
PTF11izt                & --               & --              & $4.692\pm0.104$ & --              & $ 4.692\pm 0.104$ \\
2012A                   & $ 4.854\pm0.131$ & $4.937\pm0.073$ & $4.634\pm0.086$ & $4.769\pm0.091$ & $ 4.917\pm 0.064$ \\
2012aw                  & $ 4.863\pm0.138$ & $4.889\pm0.057$ & $4.884\pm0.072$ & $4.863\pm0.068$ & $ 4.886\pm 0.052$ \\
2012br                  & --               & --              & $4.909\pm0.144$ & --              & $ 4.909\pm 0.144$ \\
2012cd                  & --               & --              & $5.000\pm0.107$ & --              & $ 5.000\pm 0.107$ \\
2012ec$^{\dagger}$      & $ 5.053\pm0.141$ & $4.721\pm0.071$ & $4.745\pm0.088$ & $4.741\pm0.092$ & $ 4.788\pm 0.063$ \\
PTF12grj                & --               & --              & $4.735\pm0.107$ & --              & $ 4.735\pm 0.107$ \\
PTF12hsx                & --               & --              & $4.974\pm0.131$ & --              & $ 4.974\pm 0.131$ \\
2013K                   & --               & $4.770\pm0.103$ & $4.721\pm0.090$ & $4.831\pm0.093$ & $ 4.721\pm 0.090$ \\
2013ab                  & --               & --              & $4.777\pm0.081$ & $4.708\pm0.082$ & $ 4.777\pm 0.081$ \\
2013am$^{\dagger}$      & --               & $4.468\pm0.079$ & $4.670\pm0.105$ & $4.533\pm0.105$ & $ 4.468\pm 0.079$ \\
2013bu                  & --               & --              & $4.573\pm0.081$ & $4.685\pm0.079$ & $ 4.685\pm 0.079$ \\
2013by                  & --               & $4.874\pm0.087$ & $4.762\pm0.101$ & $4.995\pm0.121$ & $ 4.874\pm 0.087$ \\
2013ej                  & $ 4.722\pm0.091$ & $4.736\pm0.058$ & $4.742\pm0.073$ & $4.888\pm0.070$ & $ 4.732\pm 0.049$ \\
2013fs                  & --               & --              & $4.857\pm0.085$ & $4.837\pm0.079$ & $ 4.837\pm 0.079$ \\
2013hj                  & --               & --              & $4.973\pm0.075$ & $5.035\pm0.074$ & $ 5.035\pm 0.074$ \\
iPTF13dkz               & --               & --              & $4.932\pm0.122$ & --              & $ 4.932\pm 0.122$ \\
LSQ13dpa                & --               & --              & $4.968\pm0.095$ & $4.943\pm0.128$ & $ 4.968\pm 0.095$ \\
2014G                   & --               & $4.699\pm0.076$ & $4.946\pm0.076$ & $4.949\pm0.077$ & $ 4.699\pm 0.076$ \\
2014cx                  & --               & $4.914\pm0.068$ & $4.893\pm0.084$ & $4.820\pm0.087$ & $ 4.914\pm 0.068$ \\
2014cy                  & --               & --              & $4.607\pm0.088$ & $4.560\pm0.088$ & $ 4.560\pm 0.088$ \\
ASASSN-14dq             & --               & --              & $4.864\pm0.083$ & $4.892\pm0.085$ & $ 4.864\pm 0.083$ \\
2014dw                  & --               & --              & $4.746\pm0.103$ & $4.717\pm0.131$ & $ 4.746\pm 0.103$ \\
ASASSN-14ha$^{\dagger}$ & --               & $4.433\pm0.078$ & $4.502\pm0.089$ & $4.651\pm0.107$ & $ 4.433\pm 0.078$ \\
OGLE14-18               & --               & --              & $4.835\pm0.090$ & --              & $ 4.835\pm 0.090$ \\
2015V                   & --               & --              & $4.696\pm0.085$ & $4.600\pm0.091$ & $ 4.696\pm 0.085$ \\
2015W                   & --               & --              & $4.832\pm0.086$ & $4.931\pm0.096$ & $ 4.832\pm 0.086$ \\
2015an                  & --               & --              & $4.730\pm0.087$ & $4.944\pm0.106$ & $ 4.730\pm 0.087$ \\
2015ba                  & --               & $4.442\pm0.074$ & $4.597\pm0.087$ & $4.905\pm0.080$ & $ 4.442\pm 0.074$ \\
2015bs                  & --               & $5.063\pm0.077$ & $4.941\pm0.089$ & $4.897\pm0.098$ & $ 5.063\pm 0.077$ \\
2015cz                  & --               & --              & $4.846\pm0.081$ & $4.938\pm0.089$ & $ 4.846\pm 0.081$ \\
ASASSN-15oz             & --               & $4.877\pm0.071$ & $5.082\pm0.091$ & $5.064\pm0.097$ & $ 4.877\pm 0.071$ \\
2016X                   & --               & --              & $4.844\pm0.099$ & $4.865\pm0.107$ & $ 4.844\pm 0.099$ \\
2016aqf$^{\dagger}$     & --               & $4.625\pm0.076$ & $4.683\pm0.095$ & $4.666\pm0.112$ & $ 4.625\pm 0.076$ \\
2016bkv$^{\dagger}$     & --               & --              & $4.609\pm0.085$ & $4.309\pm0.094$ & $ 4.609\pm 0.085$ \\
2016gfy                 & --               & $4.827\pm0.068$ & $4.958\pm0.084$ & $4.935\pm0.087$ & $ 4.827\pm 0.068$ \\
2016ija                 & --               & --              & $4.727\pm0.098$ & $4.826\pm0.132$ & $ 4.727\pm 0.098$ \\
2017it                  & --               & --              & $4.890\pm0.073$ & $4.856\pm0.070$ & $ 4.856\pm 0.070$ \\
2017ahn                 & --               & --              & $4.860\pm0.094$ & $4.840\pm0.126$ & $ 4.860\pm 0.094$ \\
2017eaw                 & $ 4.970\pm0.056$ & $4.933\pm0.061$ & $4.975\pm0.075$ & $4.964\pm0.077$ & $ 4.953\pm 0.041$ \\
2017gmr                 & --               & --              & $5.091\pm0.081$ & $5.057\pm0.082$ & $ 5.091\pm 0.081$ \\
2018aoq$^{\dagger}$     & $ 4.541\pm0.052$ & --              & $4.506\pm0.072$ & $4.567\pm0.068$ & $ 4.551\pm 0.041$ \\
2018cuf                 & --               & $4.859\pm0.077$ & $4.810\pm0.090$ & $4.876\pm0.105$ & $ 4.859\pm 0.077$ \\
2018hwm                 & --               & $4.361\pm0.075$ & $4.414\pm0.091$ & $4.430\pm0.091$ & $ 4.361\pm 0.075$ 
\\
\hline
\multicolumn{6}{b{0.80\textwidth}}{\textit{Notes}. $\Lf$ values are in units of $\lsun$, and quoted uncertainties are $1\,\sigma$ errors. Errors on $\log \Lf(\loi)$, $\log \Lf(\mni)$, $\log \Lf(\mv)$, and $\log\Lf$ are random and do not include the systematic calibration error of 0.043\,dex.}\\
\multicolumn{6}{l}{$^\dagger$SNe in the VL sample with $M_R^\text{max}>-16.8$.}
\end{tabular}
\end{table*}

\subsection{Selection bias correction}\label{sec:SBC}
\begin{figure}
\includegraphics[width=1.0\columnwidth]{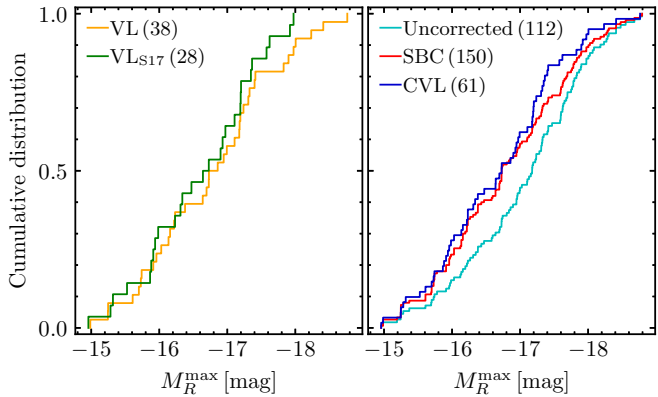}
\caption{Left-hand panel: cumulative distributions for the $M_R^\text{max}$ values in the volume-limited (orange line) and VL$_\text{S17}$ (green line) samples. Right-hand panel: cumulative $M_R^\text{max}$ distributions for the SN sample uncorrected (cyan line) and corrected for selection bias (red line), and for the combined volume-limited sample (blue line). Numbers in parentheses are the sample sizes.}
\label{fig:MRmax_cdf}
\end{figure}

The SN sample used in this work, like that of \citet{2021MNRAS.505.1742R}, is affected by selection bias. To correct for this bias, I proceed in a similar way as in \citet{2021MNRAS.505.1742R}, where volume-limited samples were used as references to compute the selection bias correction. I select the 38~SNe in my sample with ${\mu\leq31.2}$ (hereafter the VL sample) and the 28~SNe~II in the sample of \citet{2017PASP..129e4201S} with $\mu\leq32.9$ (the $\mathrm{VL_{S17}}$ sample, available in \citealt{2021MNRAS.505.1742R}). The $\mathrm{VL_{S17}}$ set has completeness $\gtrsim95$~per~cent, while SNe~II with low luminosity or high reddening have completeness $\approx70$~per~cent. To compare the VL set with the $\mathrm{VL_{S17}}$ sample, I use their $M_R^\text{max}$ values. The cumulative $M_R^\text{max}$ distributions for the two samples are shown in the left-hand panel of Fig.~\ref{fig:MRmax_cdf}. I use the $k$-sample Anderson-Darling (AD) test \citep{Scholz_Stephen1987} to evaluate whether the $M_R^\text{max}$ samples are drawn from a common unspecified distribution (the null hypothesis), obtaining a standardized test statistics ($T_\text{AD}$) of $-0.73$ with a $p$-value of 0.80. This means that the null hypothesis cannot be rejected at the 80~per~cent significance level. Given that the $\mathrm{VL_{S17}}$ and the VL samples are likely drawn from a common $M_R^\text{max}$ distribution, I combine them into a single data set, which I refer to as the combined volume-limited (CVL) sample.

Correcting the SN sample for selection bias is equivalent to correcting the set of 74~SNe that are not in the VL sample, which I refer to as the non-complete (NC) set. Table~\ref{table:completeness} lists the number of SNe in four $M_R^\text{max}$ bins of width 1\,mag (Column~1) for the NC set (Column~2) and the CVL sample (Column~3). The NC sample is almost the same as that of \citet{2021MNRAS.505.1742R}, for which the selection bias correction is practically negligible for $M_R^\text{max}<-16.8$\,mag. Therefore, I assume that the NC sample is complete at least down to $M_R^\text{max}=-16.8$\,mag. Under this assumption, I scale the number of SNe in the CVL sample by a factor of 1.83 to match the number of SNe with $M_R^\text{max}<-16.8$ in the NC and CVL samples, and then subtract the number of SNe in the NC sample. The resulting numbers, listed in Column~4, correspond to the selection bias correction. This is practically zero for the two brightest $M_R^\text{max}$ bins, and virtually twice the number of SNe in the VL sample with $M_R^\text{max}>-16.8$ (see Column~5). Therefore, to correct the SN sample used in this work for selection bias, I include twice the SNe in the VL sample with $M_R^\text{max}>-16.8$. Those SNe are marked with a dagger in Table~\ref{table:Lf}. I refer to this sample of 150~SNe as the selection bias corrected (SBC) sample, which is virtually complete except for SNe~II with low luminosity or high reddening.

\begin{table}
\caption{Histograms for $M_R^{\max}$.}
\label{table:completeness}
\begin{tabular}{ccccc}
\hline
 $M_R^{\max}$ range & NC & CVL & $1.83\,\text{CVL}-\text{NC}$ & VL \\
\hline
 $-17.8,\,-18.8$    & 16 &   8 & $-1$                         &  7 \\ 
 $-16.8,\,-17.8$    & 37 &  21 &    1                         & 12 \\  
 $-15.8,\,-16.8$    & 15 &  21 &   23                         & 12 \\  
 $-14.8,\,-15.8$    &  6 &  11 &   14                         &  7 \\ 
\hline
\end{tabular}
\end{table}

The right-hand panel of Fig.~\ref{fig:MRmax_cdf} shows the cumulative distributions for the $M_R^\text{max}$ values in the sample used in this work, uncorrected and corrected for selection bias. For comparison, I include the distribution for the CVL set, where its similarity to the SBC sample is evident.

\subsection{Linear regression}\label{sec:posterior_prob_maximization}
As we will see in the next section, correlations between $\log\Lf$ and the SN observables $\log\loi$, $\log\mni$, and $\mv$ can be expressed as linear regressions. Let $\bar{y}(\theta,x)$ be the model that describes the linear correlation between the observables $x$ and $y$, where $\theta$ is a vector containing the free parameters of the model. Given $n$ measurements of $x$, $y$, and their $1\,\sigma$ errors ($\sigma_x,\sigma_y$), I compute $\theta$ by maximizing the posterior probability
\begin{equation}\label{eq:posterior_prob}
p(\theta|x,y,\sigma_x,\sigma_y)=p(\theta)\mathcal{L}(\theta|x,y,\sigma_x,\sigma_y),
\end{equation}
where $p(\theta)$ is the prior function (assumed to be uninformative in this work), and $\mathcal{L}(\theta|x,y,\sigma_x,\sigma_y)$ is the likelihood of the linear model, given by
\begin{equation}
\ln\mathcal{L}(\theta|x,y,\sigma_x,\sigma_y)\propto\sum_{i=1}^n \left[\ln V_i+\frac{(y_i-\bar{y}(\theta,x_i))^2}{V_i}\right].
\end{equation}
Here, $V_i=  \sigma_{y_i}^2+(\partial\bar{y}/\partial x)^2\sigma_{x_i}^2-2(\partial\bar{y}/\partial x)\sigma_{xy,i}+\sigma_0^2$
is the variance of $y_i-\bar{y}(\theta,x_i)$, where $\sigma_0$ is the error not accounted for the errors in $x_i$ and $y_i$, and
\begin{equation}
\sigma_{xy,i}=\frac{\partial x_i}{\partial \mu}\frac{\partial y_i}{\partial \mu}\sigma_\mu^2 + \frac{\partial x_i}{\partial E_\bv}\frac{\partial y_i}{\partial E_\bv}\sigma_{E_\bv}^2
\end{equation}
is the covariance between $x_i$ and $y_i$. I maximize the posterior probability in equation~(\ref{eq:posterior_prob}) by means of a Markov Chain Monte Carlo process using the python package \texttt{emcee} \citep{2013PASP..125..306F}, which also provides the marginalized distributions of the parameters. I adopt the $\ssd$ values of those distributions as the $1\,\sigma$ errors of the free parameters.

\section{Results}\label{sec:results}

\subsection{Progenitor luminosities}

\subsubsection{Empirical correlations}\label{sec:empirical_correlations}
\begin{figure}
\includegraphics[width=1.0\columnwidth]{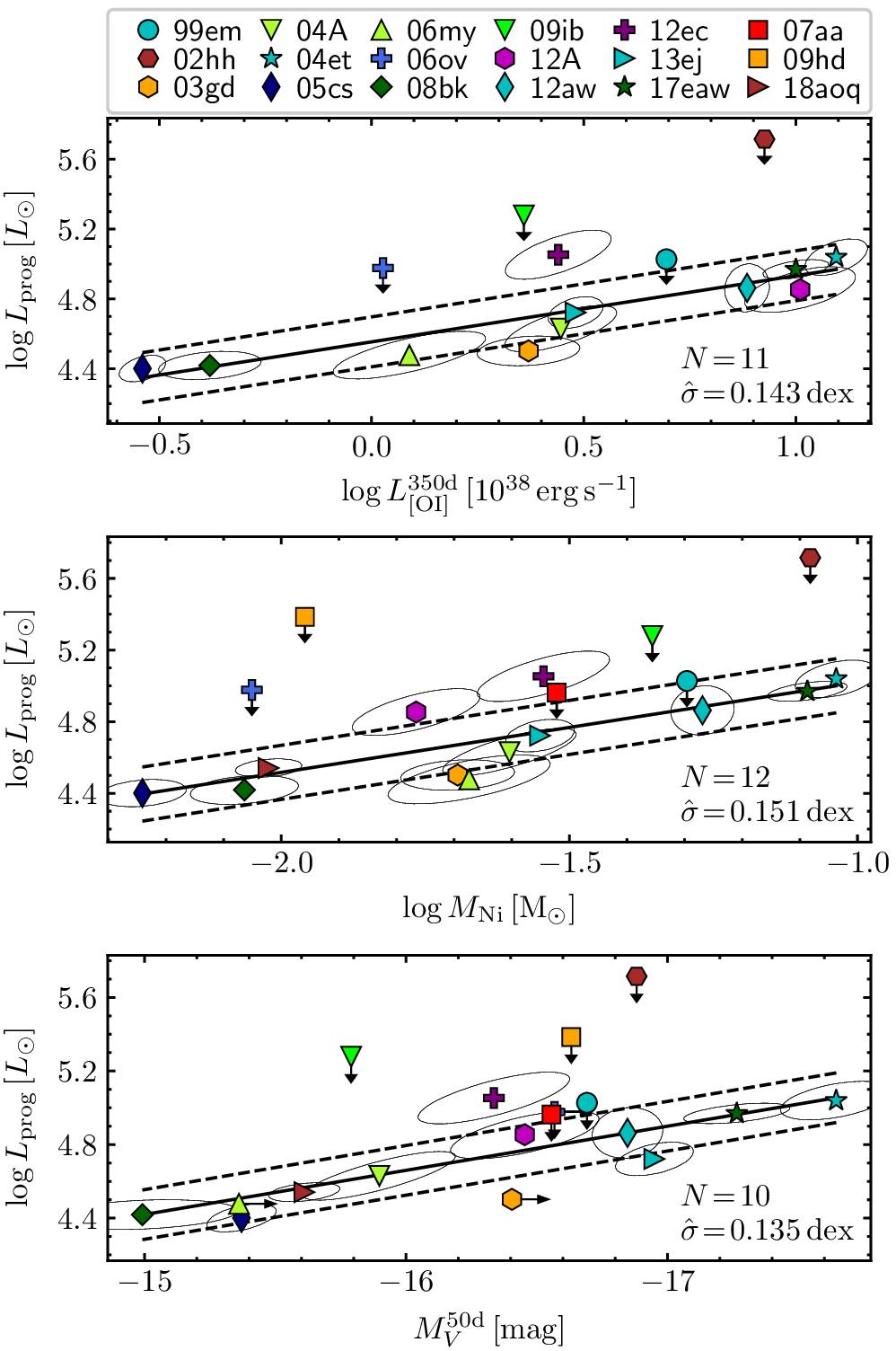}
\caption{Progenitor luminosity against $\loi$ (top panel), $^{56}$Ni mass (middle panel), and $\mv$ (bottom panel). Solid lines are straight line fits and dashed lines are $\pm1\,\ssd$ limits. Ellipses are $1\,\sigma$ confidence regions, while arrows indicate upper limits.}
\label{fig:Lf_correlations}
\end{figure}

Fig.~\ref{fig:Lf_correlations} shows $\log\Lf$ against $\log\loi$ (top panel), $\log\mni$ (middle panel), and $\mv$ (bottom panel). For these pairs of variables I compute absolute Pearson correlation coefficients $|r_\text{p}|$ of 0.84, 0.81, and 0.86, respectively, indicating strong linear correlations. The probability of obtaining $|r_\text{p}|$ values $\geq0.84$, 0.81, and 0.86 from random populations of size 11, 12, and 10, respectively, is of 0.1~per~cent. The resulting correlation between $\log\Lf$ and $\log\mni$ confirms the finding of \citet{2011MNRAS.417.1417F} and \citet{2015arXiv150602655K}.

I use the procedure described in Section~\ref{sec:posterior_prob_maximization} to compute the parameters of the linear correlations mentioned above, obtaining
\begin{equation}\label{eq:Lf_LOI}
\log\Lf = 4.554(42)+0.379(58)\log\loi
\end{equation}
($\Lf$ in $\lsun$ and $\loi$ in $10^{38}$\,erg\,s$^{-1}$) with $\sigma_0=0.053$ and $\ssd=0.143$, valid in the range $-0.54\leq\log\loi\leq 1.09$,
\begin{equation}\label{eq:Lf_MNi}
\log\Lf = 5.521(147)+0.502(87)\log\mni
\end{equation}
($\mni$ in $\msun$) with $\sigma_0=0.066$ and $\ssd=0.151$, valid in the range $-2.24\leq\log\mni\leq-1.04$, and
\begin{equation}\label{eq:Lf_MV50d}
\log\Lf = 0.820(692)-0.240(42) \mv
\end{equation}
with $\sigma_0=0.060$ and $\ssd=0.135$, valid in the range ${-17.7\leq\mv\leq-15.0}$. In these expressions, numbers in parentheses are $1\,\sigma$ errors in units of 0.001.

\begin{figure}
\includegraphics[width=1.0\columnwidth]{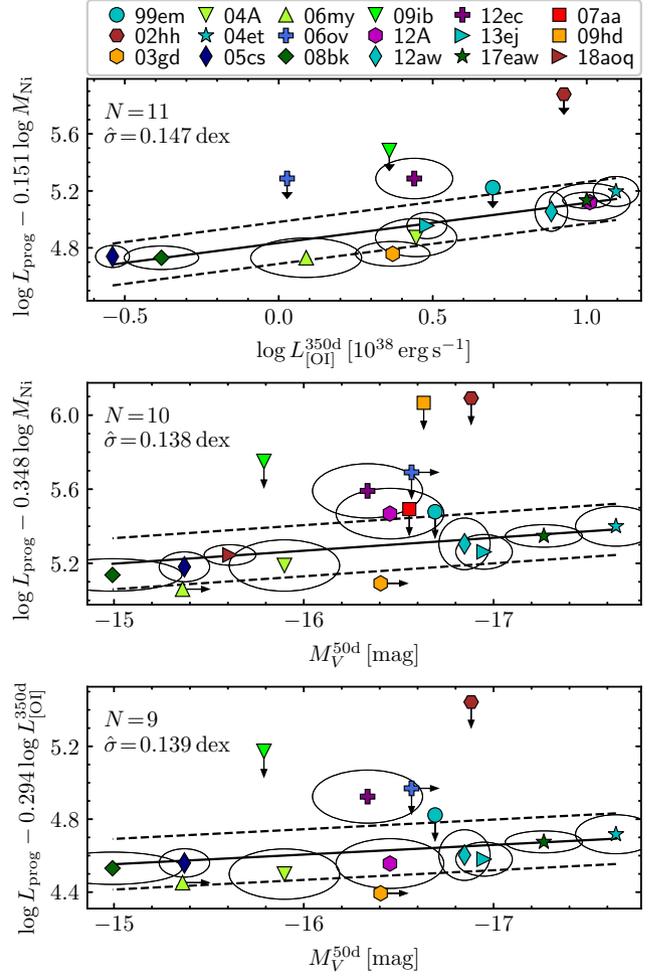}
\caption{Progenitor luminosity against $\mni$ and $\loi$ (top panel), $\mni$ and $\mv$ (middle panel), and $\loi$ and $\mv$ (bottom panel). Lines, ellipses, and arrows have the same meaning as in Fig.~\ref{fig:Lf_correlations}.}
\label{fig:Lf_l_3d}
\end{figure}

I also compute the linear dependence of $\log\Lf$ on two SN~II observables, obtaining
\begin{equation}
\log\Lf = 4.835+0.281\log\loi+0.151\log\mni
\end{equation}
with $\sigma_0=0.057$\,dex and ${\ssd=0.147}$\,dex,
\begin{equation}
\log\Lf= 4.148 + 0.348\log\mni-0.070 \mv
\end{equation}
with $\sigma_0=0.006$\,dex and $\ssd=0.138$\,dex, and
\begin{equation}
\log\Lf= 3.758 -0.053 \mv + 0.294\log\loi
\end{equation}
with $\sigma_0=0.004$\,dex and $\ssd=0.139$\,dex. Those correlations are shown in Fig.~\ref{fig:Lf_l_3d}. The $\ssd$ values do not decrease by adding a third variable to the correlation, which means that the inclusion of that variable does not provide further information  about $\Lf$. This is probably because $\mv$ is correlated with $\mni$ \citep[e.g.][]{2003ApJ...582..905H,2014MNRAS.439.2873S,2016MNRAS.459.3939V,2021MNRAS.505.1742R}, while the luminosity of the [\ion{O}{I}] doublet depends not only on oxygen mass but also on $\mni$ \citep[e.g.][]{2003MNRAS.338..939E}.

\subsubsection{Progenitor luminosity estimates}
Equations~(\ref{eq:Lf_LOI}), (\ref{eq:Lf_MNi}), and (\ref{eq:Lf_MV50d}) allow us to compute $\log\Lf$ from $\log\loi$, $\log\mni$, and $\mv$, respectively. The random error on the measured $\log\Lf$ is given by
\begin{equation}
\sigma_{\log\Lf}=\sqrt{(\partial\log\Lf/\partial x)^2\sigma_{x}^2+\sigma_0^2},
\end{equation}
where $x=\log\loi$, $\log\mni$, or $\mv$, while the systematic uncertainty on $\log\Lf$ due to the calibration error is of $\ssd/\sqrt{N}=0.043$\,dex.

I compute $\log\Lf$ for the 44~SNe with $\log\loi$ measurements ($\log\Lf(\loi)$), the 111~SNe with $\log\mni$ estimates ($\log \Lf(\mni)$), and the 93~SNe with $\mv$ values ($\log\Lf(\mv)$). Those estimates are listed in Columns~3, 4, and 5 of Table~\ref{table:Lf}, respectively. Of the reported $\log\Lf$ values, 14 are computed by extrapolating equations~(\ref{eq:Lf_LOI})--(\ref{eq:Lf_MV50d}). Specifically, SNe~1992H, 1996W, 2007it, and 2015bs have $\log\loi>1.09$\,dex; SNe~1992H, 1996W, 2007it, 2009ay, ASASSN-15oz, and 2017gmr have $\log\mni>-1.04$\,dex; and SNe~1994N, 2003B, 2003Z, and 2016bkv have $\mv>-15$\,mag. For these SNe, the mean offsets of their extrapolated $\log\Lf(\loi)$, $\log \Lf(\mni)$, and $\log\Lf(\mv)$ values with respect to the $\log\Lf$ estimates computed with parameters lying within the validity ranges are of $0.03\pm0.07$, $0.03\pm0.03$, and $0.10\pm0.14$\,dex, respectively. This indicates that the extrapolation of equations~(\ref{eq:Lf_LOI})--(\ref{eq:Lf_MV50d}) for the SNe mentioned above provides appropriate $\log\Lf$ values.

The mean errors on $\log\Lf(\loi)$, $\log\Lf(\mni)$, and $\log\Lf(\mv)$ are of 0.086, 0.101, and 0.105\,dex, respectively, corresponding to $\Lf$ errors of 20, 23, and 24~per~cent, respectively. The calibration error and $\sigma_0$ induce 67, 64, and 55~per~cent of the total error on $\log\Lf(\loi)$, $\log\Lf(\mni)$, and $\log\Lf(\mv)$, respectively.

\begin{figure}
\includegraphics[width=1.0\columnwidth]{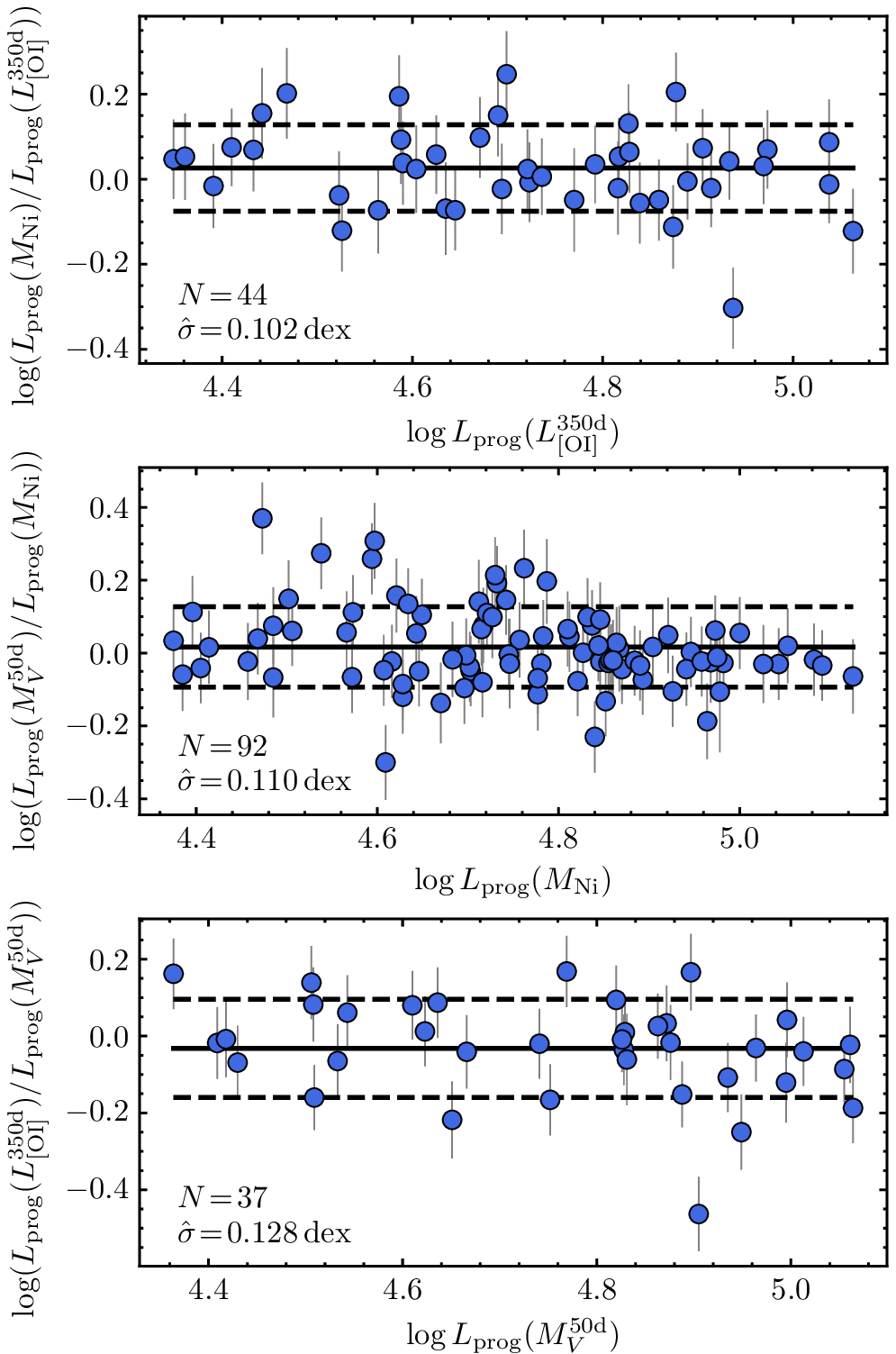}
\caption{Differences between $\log\Lf$ values computed with $\mni$ and $\loi$ (top panel), $\mv$ and $\mni$ (middle panel), and $\loi$ and $\mv$ (bottom panel). Solid lines indicate mean values, while dashed lines are the $\pm1\,\ssd$ limits around the means. Error bars are $1\,\sigma$ errors.}
\label{fig:lLf_comparison}
\end{figure}

Fig.~\ref{fig:lLf_comparison} shows the differences between $\log\Lf(\mni)$ and $\log\Lf(\loi)$ (top panel), $\log\Lf(\mv)$ and $\log\Lf(\mni)$ (middle panel), and $\log\Lf(\loi)$ and $\log\Lf(\mv)$ (bottom panel). For those differences I compute mean offsets of $0.026$, $0.017$, and $-0.032$\,dex, respectively, and $\ssd$ values of 0.102, 0.110, and 0.128\,dex, respectively. These offsets are consistent with zero within $1.7\,\ssd/\sqrt{N}$, meaning that the $\log\Lf$ values computed from $\log\loi$, $\log\mni$, and $\mv$ are statistically consistent. Therefore, I adopt the $\log\Lf$ value with the lowest uncertainty as the best $\log\Lf$ estimate. If $\log\Lf(\text{phot})$ is available, then I adopt the weighted average between this value and the $\log\Lf$ estimate with the lowest error as the best $\log\Lf$ value. These estimates are listed in Column~6 of Table~\ref{table:Lf}.

\subsection{Progenitor luminosity distribution}
\begin{figure}
\includegraphics[width=1.0\columnwidth]{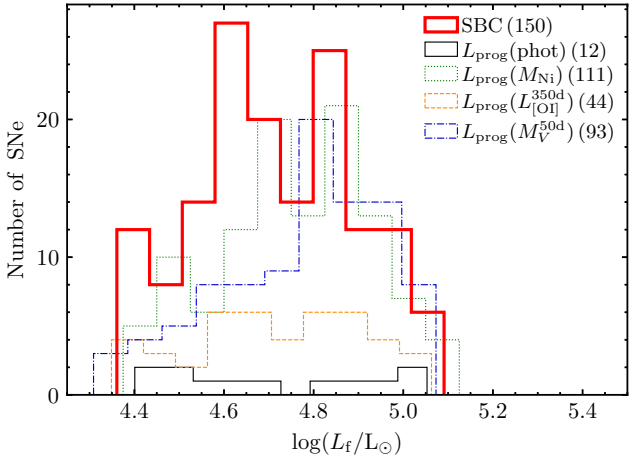}
\caption{Histogram for the final luminosities in the selection-bias corrected sample (red thick line), along with the histograms for the $\log\Lf$ values computed with progenitor photometry (black thin line), $\mni$ (green dotted line), $\loi$ (orange dashed line), and $\mv$ (blue dash-dotted line). Numbers in parentheses are the sample sizes.}
\label{fig:lLf_histogram}
\end{figure}

Fig.~\ref{fig:lLf_histogram} shows the progenitor luminosity distribution for the SBC sample, along with the distributions for the $\log\Lf(\text{phot})$, $\log\Lf(\mni)$, $\log\Lf(\loi)$, and $\log\Lf(\mv)$ values. The mean, $\ssd$, minimum, and maximum values of these distributions are summarized in Table~\ref{table:lLf_stats}. The minimum and maximum values of the distributions for $\log\Lf(\mni)$, $\log\Lf(\loi)$, and $\log\Lf(\mv)$ are statistically consistent with those of the $\log\Lf(\text{phot})$ distribution. In other words, the SNe in this work with the lowest (largest) $^{56}$Ni mass, the lowest (highest) $\log\loi$ estimate, and the highest (lowest) $\mv$ value have progenitors luminosities consistent with the luminosity of the faintest (brightest) progenitor detected in pre-explosion images. The apparent upper limit of $\log(L/\lsun)\approx5.1$ for the progenitor luminosity is consistent with that found by \citet{2015PASA...32...16S}.

\begin{table}
\caption{Statistics of the $\log (\Lf/\lsun)$ distributions.}
\label{table:lLf_stats}
\begin{tabular}{lccccc}
\hline
Sample             & $N$   & Mean    & $\ssd$  & Min     & Max     \\
\hline    
SBC                & $150$ & $4.716$ & $0.18$ & $4.361( 75)$ & $5.091( 81)$ \\  
$\Lf(\text{phot})$ & $ 12$ & $4.706$ & $0.24$ & $4.401( 76)$ & $5.053(141)$ \\  
$\Lf(\mni)$        & $111$ & $4.757$ & $0.18$ & $4.375( 85)$ & $5.125(111)$ \\  
$\Lf(\loi)$        & $ 44$ & $4.719$ & $0.20$ & $4.349( 57)$ & $5.063( 77)$ \\  
$\Lf(\mv)$         & $ 92$ & $4.771$ & $0.18$ & $4.309( 94)$ & $5.073( 95)$ \\
\hline
\multicolumn{6}{b{0.95\columnwidth}}{\textit{Note}: Numbers in parentheses are $1\,\sigma$ random errors in units of 0.001.}
\end{tabular}
\end{table}

\subsection{Comparison with RSG luminosities}\label{sec:comparison_with_RSGs}
For the comparison between luminosities of SN~II progenitors in the SBC sample and of RSGs in the samples of LMC, SMC, M31 and M33, I select RSGs with $\log(L/\lsun)\geq4.361$, corresponding to the minimum $\log(L/\lsun)$ value in the SBC sample.

\subsubsection{Luminosity distributions}
\begin{figure}
\includegraphics[width=1.0\columnwidth]{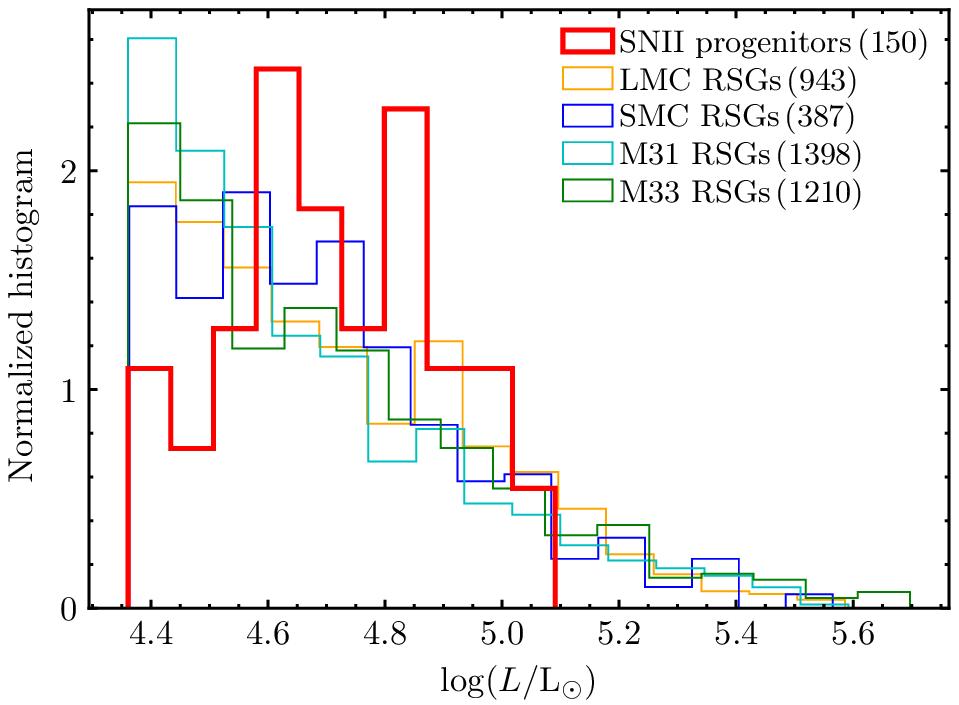}
\includegraphics[width=1.0\columnwidth]{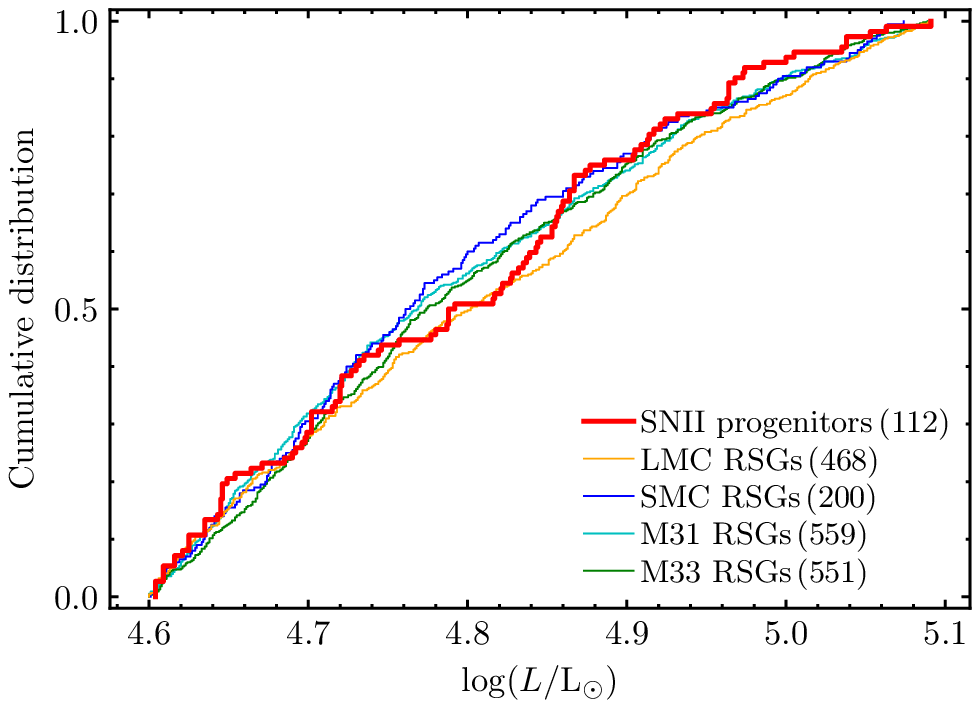}
\caption{Top panel: histograms for the progenitor luminosities in the SBC sample (thick red line) and the luminosities of the RSGs in LMC, SMC, M31, and M33 with $\log(L/\lsun)\geq4.361$ (thin lines). Bottom panel: cumulative distributions for the progenitor luminosities in the gold sample (thick red line) and the luminosities in the RSG samples in the luminosity domain of the gold sample (thin lines). Numbers in parentheses are the sample sizes.}
\label{fig:SBC_RSG_histogram}
\end{figure}

The top panel of Fig.~\ref{fig:SBC_RSG_histogram} shows the distribution for the progenitor luminosities in the SBC sample and the luminosity distributions for the RSGs in LMC, SMC, M31, and M33. We see a conspicuous absence of SN~II progenitors with ${\log(L/\lsun)>5.1}$ with respect to the RSG samples, which will be analysed in Section~\ref{sec:RSG_problem}. We also see that for $\log(L/\lsun)<4.6$ the SBC sample has a lower number density than the RSGs sets. The latter is most likely due to the SBC sample is not complete for low luminosity SNe~II. Indeed, using equation~(\ref{eq:Lf_MV50d}), progenitors with ${\log(\Lf/\lsun)<4.6}$ correspond to SNe~II with ${\mv>-15.8}$, which characterizes the population of low luminosity SNe~II (see e.g. Fig.~11 of \citealt{2021AA...655A..90Y}). In order to avoid the low completeness of low luminosity SNe~II, I select from the SBC sample those SNe with $\log(\Lf/\lsun)\geq4.6$, which I refer to as the gold sample. Given the high completeness of the SNe that are not low luminosity SNe~II (see Section~\ref{sec:SBC}), I consider the gold sample to be complete to $\log(L/\lsun)=4.6$\,dex.

The bottom panel of Fig.~\ref{fig:SBC_RSG_histogram} shows the cumulative luminosity distributions for the gold sample and the RSGs in LMC, SMC, M31, and M33 in the luminosity domain of the gold sample ($4.6\leq\log(L/\lsun)\leq5.091$). Using the $k$-sample AD test to compare the gold sample with these RSG sets, I obtain $T_\text{AD}$ ($p$-value) of $0.22$ (0.28), $-0.39$ (0.55), $-0.52$ (0.64), and $-0.29$ (0.50), respectively. Therefore, the null hypothesis that the SN~II progenitors in the gold sample and RSGs with $4.6\leq\log(L/\lsun)\leq5.091$ are drawn from a common luminosity distribution cannot be rejected at a significance level of at least 28~per~cent. This result supports RSGs as SN II progenitors.

\subsubsection{The RSG problem}\label{sec:RSG_problem}
\begin{figure}
\includegraphics[width=1.0\columnwidth]{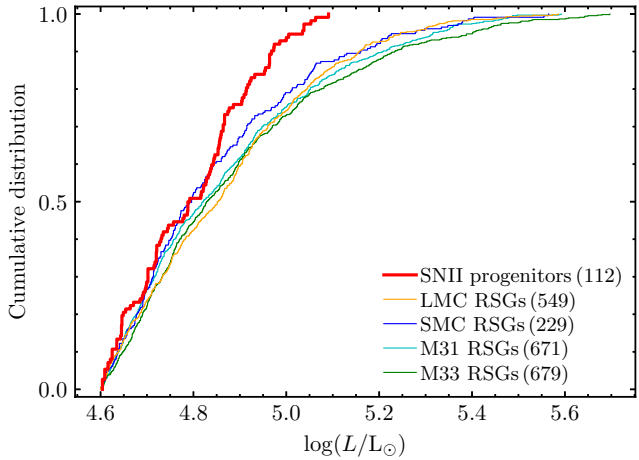}
\caption{Cumulative distributions for the progenitor luminosities in the gold sample (red thick line) and the luminosities of the RSGs in LMC, SMC, M31, and M33 with ${\log(L/\lsun)\geq4.6}$ (thin lines). Numbers in parentheses are the sample sizes.}
\label{fig:SNe_RSG_histogram}
\end{figure}

Fig.~\ref{fig:SNe_RSG_histogram} shows the cumulative luminosity distributions for the gold sample and RSGs in LMC, SMC, M31, and M33 with $\log(L/\lsun)\geq4.6$. As previously mentioned, there is a conspicuous absence of SN~II progenitors with $\log(L/\lsun)>5.1$ with respect to the RSG samples. For those samples, between 13 and 18~per~cent of the RSGs with $\log(L/\lsun)\geq4.6$ have $\log(L/\lsun)>5.1$. If such RSGs explode as SNe~II, then the gold sample should have between 16 and 25 progenitors with $\log(L/\lsun)>5.1$. 
The Poissonian probability of not observing such events is given by
\begin{equation}\label{eq:Poisson}
P=\frac{e^{-n_\text{exp}}n_\text{exp}^{n_\text{obs}}}{n_\text{obs}!},
\end{equation}
where $n_\text{exp}$ and $n_\text{obs}$ are the expected and observed number of progenitors with $\log(L/\lsun)>5.1$, respectively. For $n_\text{obs}=0$ and $n_\text{exp}=16$ (25), $P=1.1\times10^{-7}$ ($1.4\times10^{-11}$), corresponding to a significance of $5.3\,\sigma$ ($6.8\,\sigma$). Since the calculation does not include luminosity uncertainties, the inferred significance values for the RSG problem are overestimated. To include the effect of luminosity errors, I perform $10^4$ simulations varying randomly luminosities in the SBC and RSG samples according to their statistical errors (assumed normal). For each realization, I construct a gold sample by selecting the $\log(L/\lsun)$ values greater than 4.6\,dex from the simulated SBC sample. The $\log(L/\lsun)$ estimates in the simulated gold sample are then shifted by a constant, which is randomly selected from a normal distribution with zero mean and standard deviation equal to the systematic calibration error. For each of the four simulated RSG samples, I compute the ratio between the number of RSGs with $\log(L/\lsun)>5.1$ and with $\log(L/\lsun)$ greater than the minimum $\log(L/\lsun)$ value in the simulated gold sample. Then, I compute $n_\text{exp}$ and $n_\text{obs}$ and, using equation~(\ref{eq:Poisson}), $P$ and the corresponding significance.

\begin{figure}
\includegraphics[width=1.0\columnwidth]{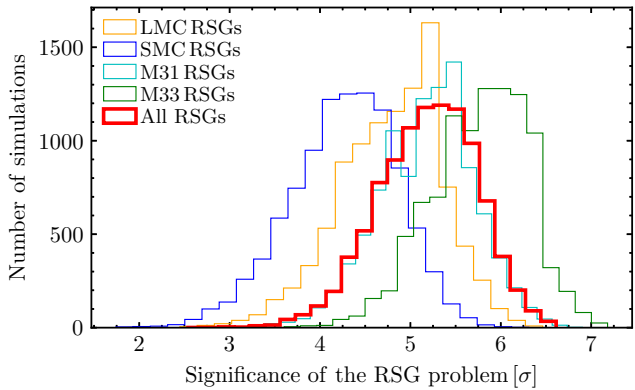}
\caption{Histograms for the significance values of the RSG problem computed with $10^4$ simulations of the gold sample, the four RSG samples (thin lines), and the combined RSG sample (thick red line).}
\label{fig:RSG_problem}
\end{figure}

Fig.~\ref{fig:RSG_problem} shows the histograms for the significance values of the RSG problem computed with simulated gold and RSG samples. For the comparison between the gold sample and the LMC, SMC, M31, and M33 sets, the mean significance values (in $\sigma$ units) are of $4.8_{-1.3}^{+1.0}$, $4.3_{-1.3}^{+1.0}$, $5.2_{-1.1}^{+1.0}$, and $5.8_{-1.1}^{+1.0}$ (95~per~cent confidence interval), respectively. Combining the four RSG samples into a single data set and performing the simulation described earlier, I obtain a mean significance of $5.2\pm0.5\,\sigma$ ($1\,\ssd$ error). Therefore, the RSG problem is statistically significant.

\section{Discussion}\label{sec:discussion}

\subsection{Comparison with initial masses from models}\label{sec:comparison_with_other_works}
I now compare the progenitor luminosities calculated in this work with initial masses reported in the literature computed with models. For this comparison, progenitor luminosities have to be transformed to initial masses using an initial mass-final luminosity relation (MLR). This relation depends on the stellar evolution model adopted in each work.

\citet{2018ApJ...858...15M} and \citet{2020AA...642A.143M,2022AA...660A..41M} reported $M_\text{i}$ values inferred by fitting hydrodynamical models to SN data. \citet{2018ApJ...858...15M} used multiband light-curve models generated with the \texttt{SNEC} code \citep{2015ApJ...814...63M}, while \citet{2020AA...642A.143M,2022AA...660A..41M} used bolometric light curves and expansion velocity curves calculated with the model of \citet{2011ApJ...729...61B}. In addition, 19~SNe used in the present study have $M_\text{i}$ estimates computed by comparing late-time spectra with spectral models generated with the \texttt{SUMO} code \citep{2011AA...530A..45J,2012AA...546A..28J}. These SNe, along with the reported $\mu$, $E_\bv^\text{host}$, and $M_\text{i}$ values are collected in Table~\ref{table:Mi_SUMO}. \citet{2018ApJ...858...15M}, \citet{2020AA...642A.143M,2022AA...660A..41M}, and the \texttt{SUMO} code adopted non-rotating RSGs models with solar composition and $M_\text{i}$ between 9 and $25\,\msun$ as progenitors. Specifically, \citet{2018ApJ...858...15M} and the \texttt{SUMO} code used RSG models computed with the \texttt{KEPLER} code \citep[e.g.][]{2002RvMP...74.1015W}, while \citet{2020AA...642A.143M,2022AA...660A..41M} computed RSG models using the \texttt{MESA} code \citep[e.g.][]{2016ApJS..227...22F}. Using initial masses and final luminosities for $M_\text{i}\leq 25\,\msun$ reported in \citet{2002RvMP...74.1015W} and \citet{2016ApJS..227...22F} (for non-rotating models), I derive MLRs for \texttt{KEPLER} and \texttt{MESA} codes, given by $\log (M_\text{i}/\msun)=-1.263+0.495\log(\Lf/\lsun)$ and ${\log (M_\text{i}/\msun)=-1.235+0.491\log(\Lf/\lsun)}$, respectively.

\begin{table}
\caption{SNe with initial masses estimated with SUMO models.}
\label{table:Mi_SUMO}
\begin{tabular}{lcccc}
\hline
 SN          & $\mu$   & $E_\bv^\text{host}$  & $M_\text{i}/\msun$& Reference$^\dagger$  \\
\hline
 1997D       & $30.64$ & $0.0  $ &  9       & a \\
 1999em      & $30.34$ & $0.06 $ & $13\pm1$ & b \\
 2004A       & $31.35$ & $0.0  $ & 12       & c \\
 2004et      & $28.70$ & $0.117$ & 15       & d \\
 2008bk      & $27.84$ & $0.0  $ &  9       & a \\
 2012A       & $29.96$ & $0.009$ & 15       & e \\
 2012aw      & $29.98$ & $0.046$ & 15       & f \\
 2012ec      & $31.19$ & $0.087$ & 13--15   & g \\
 2013ej      & $29.93$ & $0.0  $ & 12--15   & h \\
 2014G       & $31.94$ & $0.20 $ & 15--19   & i \\
 2014cx      & $31.74$ & $0.0  $ & 15       & j \\
 ASASSN-14dq & $33.26$ & $0.0  $ & 15       & j \\
 2015W       & $33.74$ & $0.0  $ & 15       & j \\
 2015bs      & $35.40$ & $0.0  $ & 15--25   & k \\
 ASASSN-15oz & $32.30$ & $0.0  $ & 15--19   & l \\
 2016aqf     & $30.16$ & $0.0  $ & $12\pm3$ & m \\
 2016gfy     & $32.36$ & $0.14 $ & 15       & n \\
 2017eaw     & $29.44$ & $0.0  $ & 15       & o \\
 2018cuf     & $33.11$ & $0.11 $ & 12--15   & p \\
\hline
\multicolumn{5}{b{0.95\columnwidth}}{$^\dagger$(a): \citet{2018MNRAS.475..277J}; (b): \citet{2018MNRAS.474.2116D}; (c): \citet{2017MNRAS.467..369S}; (d): \citet{2012AA...546A..28J}; (e): \citet{2013MNRAS.434.1636T}; (f): \citet{2014MNRAS.439.3694J}; (g): \citet{2015MNRAS.448.2482J}; (h): \citet{2016MNRAS.461.2003Y}; (i): \citet{2016MNRAS.462..137T}; (j): \citet{2016MNRAS.459.3939V}; (k): \citet{2018NatAs...2..574A}; (l): \citet{2019MNRAS.485.5120B}; (m): \citet{2020MNRAS.497..361M}; (n): \citet{2019ApJ...882...68S}; (o): \citet{2019ApJ...875..136V}; (p): \citet{2020ApJ...906...56D}.}\\
\end{tabular}
\end{table}

To compare the initial masses calculated with the three models mentioned above (\texttt{SNEC}, the Bersten's model, and \texttt{SUMO}) with the $\log\Lf$ values computed from $\loi$, $\mni$, or $\mv$, I first recompute $\log\Lf$ using the distances and reddenings adopted in the respective work, and then convert $\log\Lf$ to $\log M_\text{i}$ ($\log M_\text{i}(\Lf)$) using the corresponding MLR. In the case of \citet{2022AA...660A..41M}, the authors do not provide $E_\bv^\text{host}$ values. Instead, they report a variable called $scale$, equivalent to the intrinsic luminosity divided by the observed luminosity uncorrected for host galaxy extinction. The $scale$ parameter accounts for host galaxy extinction and for the difference between the adopted distance and the true distance. Since I adopt the distances of \citet{2022AA...660A..41M}, I assume that they correspond to the true values, so the host galaxy extinction affecting the bolometric light curve is $A_\text{bol}=2.5\log(scale)$. \citet{2021MNRAS.505.1742R} computed $A_\text{bol}/E_\bv=1.68$, so I adopt $E_\bv^\text{host}=1.49\log(scale)$. Among the SNe in common between the sample of \citet{2022AA...660A..41M} and the one used in this work, SNe~2004ej and 2007od have $scale$ values significantly lower than unity, which result in negative $E_\bv^\text{host}$ values of $-0.14$ and $-0.39$\,mag, respectively. The low $scale$ values for these SNe could be due to, for example, an overestimation of their distances. Since the negative $E_\bv^\text{host}$ values of SNe~2004ej and 2007od have no physical meaning, I do not include those SNe in the comparison with the model of \citet{2011ApJ...729...61B}.

\begin{figure}
\includegraphics[width=1.0\columnwidth]{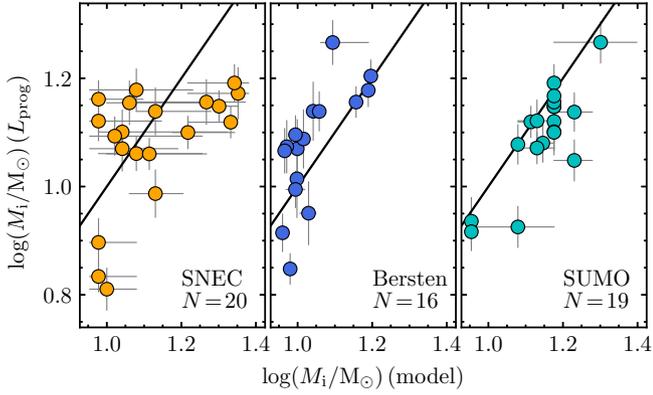}
\caption{Initial masses estimated from the recalibrated $\log\Lf$ values against those computed by \citet{2018ApJ...858...15M} using \texttt{SNEC} (left-hand panel), by \citet{2020AA...642A.143M,2022AA...660A..41M} using the model of \citet{2011ApJ...729...61B} (middle panel), and those computed with \texttt{SUMO} models (right-hand panel). Solid lines are one-to-one correspondences.}
\label{fig:lit_comp}
\end{figure}

Fig.~\ref{fig:lit_comp} shows the $\log M_\text{i}(\Lf)$ values against the initial masses of \citet{2018ApJ...858...15M} (left-hand panel), \citet{2020AA...642A.143M,2022AA...660A..41M} (middle panel), and those computed with the \texttt{SUMO} models (right-hand panel). For each comparison I fit a straight line, whose slope ($b$) is listed in Column~3 of Table~\ref{table:slopes}, and a straight line with slope of unity, whose $y$-intercept ($a$) is listed in Column~4 of Table~\ref{table:slopes}. To test if two methods of measurement are statistically consistent, it is necessary to test if $b$ and $a$ are statistically consistent with unity and zero, respectively. For this task I use the one-sample $t$-test, where the $p$-values for the null hypotheses $b=1$ and $a=0$ are listed in Columns~5 and 6 of Table~\ref{table:slopes}, respectively. I choose a significance level of 0.05 to accept the null hypothesis for $b$ and $a$. Based on this criterion, the initial masses computed with the models of \citet{2011ApJ...729...61B} and \texttt{SUMO} are statistically consistent with $\log M_\text{i}(\Lf)$. On the other hand, the $b$ value for \texttt{SNEC} is significantly lower than unity, so the initial masses reported by \citet{2018ApJ...858...15M} are not consistent with $\log M_\text{i}(\Lf)$.

\begin{table}
\caption{Slopes and $y$-intercepts of the $\log M_\text{i}$ comparisons.}
\label{table:slopes}
\begin{tabular}{lccccc}
\hline
 Method         & $N$ & $b$         & $a$          & $p_{b=1}$ & $p_{a=0}$ \\
\hline
 SNEC           & 20 & $ 0.43(17)$ & $-0.043(34)$ & $0.0$  & $0.22$ \\
 Bersten        & 16 & $ 0.98(27)$ & $ 0.035(29)$ & $0.94$ & $0.25$ \\
 SUMO           & 19 & $ 0.88(14)$ & $-0.049(24)$ & $0.41$ & $0.06$ \\
 PZ11           & 11 & $ 0.21(36)$ & $-0.159(48)$ & $0.06$ & $0.01$ \\
 CRAB           &  8 & $ 0.39(28)$ & $-0.279(48)$ & $0.07$ & $0.0$  \\
 STELLA         &  8 & $-0.50(17)$ & $-0.045(68)$ & $0.0$  & $0.53$ \\
 Age-dating     & 11 & $ 0.60(22)$ & $-0.054(35)$ & $0.10$ & $0.16$ \\

\hline
\multicolumn{6}{b{0.95\columnwidth}}{\textit{Note}: Numbers in parentheses are $1\,\sigma$ errors in units of the last significant digit.}
\end{tabular}
\end{table}

I also compare $\log M_\text{i}(\Lf)$ with initial masses computed by \citet{2017MNRAS.469.2202M} through the age-dating technique, and by other authors using three different hydrodynamical models: \citet{2017MNRAS.464.3013P} based on the model of \citet{2011ApJ...741...41P} (PZ11), \citet{2019ApJ...880...59R} using the \texttt{STELLA} code \citep{1998ApJ...496..454B}, and \citet{2019MNRAS.490.2042U} based on the \texttt{CRAB} code \citep{2004AstL...30..293U}. For these works, the MLR is not straightforward to obtain. For the sake of simplicity, I assume the average of the \texttt{KEPLER} and \texttt{MESA} MLRs, i.e., ${\log (M_\text{i}/\msun)=-1.249+0.493\log(\Lf/\lsun)}$.

\begin{figure}
\includegraphics[width=1.0\columnwidth]{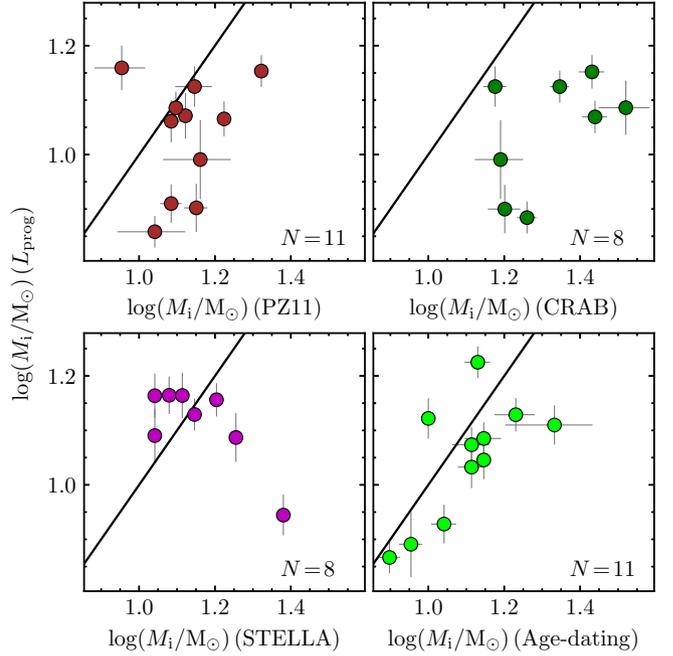}
\caption{Initial masses estimated from the recalibrated $\log\Lf$ values against those used in \citet{2017MNRAS.464.3013P} based on the model of \citet{2011ApJ...741...41P} (top left panel), in \citet{2019MNRAS.490.2042U} computed with \texttt{CRAB} (top right panel), those computed by \citet{2019ApJ...880...59R} using \texttt{STELLA} (bottom left panel), and by \citet{2017MNRAS.469.2202M} using the age-dating method (bottom right panel). Solid lines are one-to-one correspondences.}
\label{fig:lit_comp_2}
\end{figure}

Fig.~\ref{fig:lit_comp_2} shows the comparisons between $\log M_\text{i}(\Lf)$ and initial masses calculated with each of the four methods mentioned above, while the corresponding $b$, $a$, and $p$-values are listed in Table~\ref{table:slopes}. Of those methods, only the age-dating technique provides initial masses statistically consistent with $\log M_\text{i}(\Lf)$. The negative $b$ value for \texttt{STELLA} is incompatible with unity, while the $a$ values for PZ11 and \texttt{CRAB} are significantly lower than zero, so the initial masses computed with these models are not consistent with $\log M_\text{i}(\Lf)$. In particular, the $M_\text{i}$ estimates computed with the model of PZ11 and the \texttt{CRAB} code are, on average, 1.4 and 1.9 times larger than those inferred from the recalibrated $\log\Lf$ values, respectively. The overestimation of the initial masses calculated with the \texttt{CRAB} code was previously reported by \citet{2008AA...491..507U,2009AA...506..829U}.

\subsection{Systematics}

\subsubsection{BCs for RSGs}
The BC is an important source of uncertainty in determining $\log\Lf$ values. For eight of the twelve SN~II progenitors used to compute the correlations between $\log\Lf$ and SN observables, the adopted BC estimates correspond to the weighted average of the BC values for late-type RSGs ($\langle\text{BC}\rangle$). These $\langle\text{BC}\rangle$ estimates are based on only four RSGs \citep[see][]{2018MNRAS.474.2116D} so their values are not statistically robust, which could affect the inferred significance of the RSG problem. Assuming the average BC error of 0.15\,dex as the $\ssd$ around $\langle\text{BC}\rangle$, the standard error of the mean is of $\sigma_{\langle\text{BC}\rangle}=0.075$\,dex. If the true $\langle\text{BC}\rangle$ value were $3\,\sigma_{\langle\text{BC}\rangle}$ greater than the current estimate, the significance of the RSG problem would be reduced to $4.4\pm0.7\,\sigma$. This significance is greater than $3\,\sigma$ at a confidence level of 98~per~cent, so the RSG problem is still statistically significant.

\subsubsection{Calibration sample}
The correlations between $\log\Lf$ and SN observables presented in Section~\ref{sec:empirical_correlations} are based on 10--12 progenitors, so the correlation parameters could be misestimated due to the small sample sizes. Based on the observed $\ssd$ values of about 0.14--0.15\,dex and the current sample sizes, the true standard deviation around the correlations shown in Fig.~\ref{fig:Lf_correlations} ($\sigma_\text{true}$) can be as low as 0.1\,dex or as large as 0.27\,dex at a 95~per~cent confidence level.\footnote{This result is computed assuming that residuals of the correlation fits have a normal parent distribution with standard deviation $\sigma_\text{true}$, for which the quantity $(\ssd/\sigma_\text{true})^2\nu$ has a chi-square distribution with $\nu$ degrees of freedom \citep[e.g.][]{Lu_1960}.} If the $\sigma_\text{true}$ value is around 0.25\,dex, then it would be necessary to double the calibration sample size in order to have a systematic calibration error similar to the current one.

\section{Conclusion}\label{sec:conclusions}
In this work I have computed empirical correlations between luminosity of SN~II progenitors and three SN observables: $\loi$, $\mni$, and $\mv$. For this, I have used twelve SNe~II with $\Lf$ measured from progenitor photometry. Using these empirical correlations, I have estimated final luminosities for a sample of 112~SNe~II. I have corrected this sample for selection bias and, discarding low luminosity SNe~II, defined a gold sample of 112~SNe complete at $\log(L/\lsun)=4.6$\,dex.

The main conclusions are the following:
\begin{itemize}
\item[(1)] Linear correlations between $\log\Lf$ and $\log\loi$, $\log\mni$, or $\mv$ are strong and statistically significant. These correlations allow estimating $\Lf$ with a precision of 20, 23, and 24~per~cent, respectively.
\item[(2)] The luminosity distribution for the gold sample is statistically consistent with those for RSGs in SMC, LMC, M31, and M33 with $4.6\leq\log(L/L_{\sun})\leq5.091$. This reinforces the fact that SN~II progenitors correspond to RSGs.
\item[(3)] The conspicuous absence of SN~II progenitors with $\log(L/\lsun)>5.1$ with respect to what is observed in RSG luminosity distributions is significant at a $5.2\pm0.5\,\sigma$ level. This indicates that the RSG problem is statistically significant.
\item[(4)] Initial progenitor masses calculated with the hydrodynamical model of \citet{2011ApJ...729...61B}, the nebular spectra models generated with the \texttt{SUMO} code, and with the age-dating technique are statistically consistent with those computed from empirical $\log\Lf$ values and the corresponding MLR.
\end{itemize}

\section*{Acknowledgements}
I thank K. Maguire, R. Roy, F. Huang, R. Dastidar, Y. Dong, D. O'Neill, and D. Tsvetkov for sharing spectra with me. This paper is part of a project that has received funding from the European Research Council (ERC) under the European Union's Seventh Framework Programme, Grant agreement No. 833031 (PI: Dan
Maoz). This work has made use of the Weizmann Interactive Supernova Data Repository (\url{https://www.wiserep.org}). This research has made use of the Spanish Virtual Observatory (\url{https://svo.cab.inta-csic.es}) project funded by MCIN/AEI/10.13039/501100011033/ through grant PID2020-112949GB-I00. This work is based in part on observations collected at the European Organisation for Astronomical Research in the Southern Hemisphere, Chile as part of PESSTO, (the Public ESO Spectroscopic Survey for Transient Objects Survey) ESO program 188.D-3003, 191.D-0935, 197.D-1075.

\section*{Data availability}

The data underlying this article will be shared on reasonable request to the corresponding author.

\bibliographystyle{mnras}
\bibliography{references} 



\appendix

\section{SN\lowercase{e} 2015\lowercase{bs} \lowercase{and} 2018\lowercase{aoq}}\label{sec:SNe_appendix}
Here I estimate distances, reddenings, explosion epochs, absolute magnitudes, and $^{56}$Ni masses for SNe~2015bs and 2018aoq using the same methodology as in \citet{2021MNRAS.505.1742R}. For this, I use photometric and spectroscopic data presented by \citet{2019AA...622L...1O} and \citet{2019MNRAS.487.3001T,2021AstL...47..291T} for SN~2018aoq, and by \citet{2018NatAs...2..574A} for SN~2015bs.

SN~2018aoq was discovered in NGC~4151 (${cz=997}$\,km\,s$^{-1}$, \citealt{2013MNRAS.428.1790W}) by the Lick Observatory Supernova Search on 2018 April 01.436\,UT \citep{2018ATel11498....1N}. The SN, also visible in pre-explosion images taken on March 31.962\,UT \citep{2018ATel11498....1N}, was not detected on March 31.5\,UT \citep{2018ATel11526....1Y}. Using the last non-detection and the first detection epoch, along with optical spectroscopy and the \texttt{SNII\_ETOS} code\footnote{\url{https://github.com/olrodrig/SNII\_ETOS}} \citep{2019MNRAS.483.5459R}, the explosion epoch is estimated to be MJD\,$58208.74\pm0.14$. I adopt the SN~Ia distance modulus of $30.99\pm0.06$\,mag measured by \citet{2020ApJ...902...26Y}, and a Galactic reddening of $0.023\pm0.004$\,mag \citep{2011ApJ...737..103S}. I derive a host galaxy reddening of $0.085\pm0.066$\,mag using the colour method \citep{2010ApJ...715..833O}, $0.115\pm0.058$\,mag using the colour-colour method \citep{2014AJ....148..107R,2019MNRAS.483.5459R} implemented in the \texttt{C3M} code,\footnote{\url{https://github.com/olrodrig/C3M}} and $0.0\pm0.1$\,mag using the spectrum-fitting technique \citep[e.g.][]{2010ApJ...715..833O,2021MNRAS.505.1742R}. I adopt the weighted average of these three values ($0.086\pm0.040$\,mag) as the host galaxy reddening for SN~2018aoq. I compute $M_R^\text{max}=-16.029$ and $\mv=-15.613\pm0.136$. Using the $I$-band photometry in the radioactive tail and the \texttt{SNII\_nickel} code\footnote{\url{https://github.com/olrodrig/SNII\_nickel}} \citep{2021MNRAS.505.1742R}, I measure a $\log(\mni/\msun)$ value of $-2.022\pm0.058$\,dex. This value, equivalent to $\mni=0.0096\pm0.0013\,\msun$, compares to the $^{56}$Ni mass of $0.01\,\msun$ adopted by \citet{2021AstL...47..291T} for their radiation-hydrodynamical simulations.

SN~2015bs was discovered by the Catalina Real-Time Transient Survey on 2014 September 25\,UT, being not detected ten days before the discovery \citep{2018NatAs...2..574A}. The Galactic reddening toward the SN is of ${0.044\pm0.007}$\,mag \citep{2011ApJ...737..103S}, while the heliocentric redshift is of 0.027 \citep{2018NatAs...2..574A}. Using the Hubble law with a local Hubble constant of ${74.03\pm1.42}$\,km\,s$^{-1}$\,Mpc$^{-1}$ \citep{2019ApJ...876...85R} and a velocity dispersion of 382\,km\,s$^{-1}$ to account for the effect of peculiar velocities over distances, I compute a distance modulus of $35.10\pm0.11$\,mag. I estimate the explosion epoch to be MJD\,$56921.53\pm2.57$ using the \texttt{SNII\_ETOS} code. The host galaxy reddening computed with the spectrum-fitting technique is of $0.0\pm0.1$\,mag. I calculate ${M_R^\text{max}=-17.47}$ and $\mv=-16.988\pm0.325$. To estimate $\mni$, I first convert the Pan-STARRS1 $w$-band photometry to $R$-band magnitudes. For this, I use the methodology described in \citet{2021MNRAS.505.1742R}, finding a transformation given by $w-R=0.41\pm0.04$\,mag for the radioactive tail. Using the $R$-band magnitudes in the radioactive tail and the \texttt{SNII\_nickel} code, I compute $\log(\mni/\msun)=-1.156\pm0.119$\,dex, equivalent to $\mni=0.072\pm0.020\,\msun$. This estimate is statistically consistent with the $^{56}$Ni mass of $0.049\pm0.008\msun$ reported in \citet{2018NatAs...2..574A}.


\bsp	
\label{lastpage}
\end{document}